\documentclass[longauth]{aa} 
\usepackage[colorlinks=true, linkcolor=blue, citecolor=blue, urlcolor=blue]{hyperref}
\usepackage{amsmath}
\usepackage{graphicx}
\usepackage{amssymb}
\usepackage{mathrsfs}
\usepackage{caption}
\usepackage{float}
\usepackage{afterpage}
\usepackage{amstext}
\usepackage{multirow}

\usepackage{txfonts}
\usepackage{natbib}

\begin{document}

   \title{Gaia-ESO Survey: Detailed elemental abundances in red giants of the peculiar globular cluster NGC\,1851\thanks{Based on data products from observations made with ESO Telescopes at the La Silla Paranal Observatory under programme ID 188.B-3002 (The Gaia-ESO Public Spectroscopic Survey, PIs G. Gilmore and S. Randich).}$^,$\thanks{Table~3 is only available in electronic form at the CDS via anonymous ftp to cdsarc.u-strasbg.fr (130.79.128.5) or via http://cdsweb.u-strasbg.fr/cgi-bin/qcat?J/A+A/}
}
   \author{
G. Tautvai\v{s}ien\.{e}\inst{1},      
A. Drazdauskas\inst{1},
A. Bragaglia\inst{2},
S. L. Martell\inst{3,4}, 
E. Pancino\inst{5,6},
C. Lardo\inst{7},
\v{S}. Mikolaitis\inst{1},\\
R. Minkevi\v{c}i\={u}t\.{e}\inst{1},  
E. Stonkut\.{e}\inst{1},
M. Ambrosch\inst{1},
V. Bagdonas\inst{1},
Y. Chorniy\inst{1},
N. Sanna\inst{5},
E. Franciosini\inst{5},
R. Smiljanic\inst{8},
S. Randich\inst{5},
G. Gilmore\inst{9},
T. Bensby\inst{10}
M. Bergemann\inst{11},
A. Gonneau\inst{9},
G. Guiglion\inst{12},
G. Carraro\inst{13},
U. Heiter\inst{14},\\
A. Korn\inst{14},
L. Magrini\inst{5},
L. Morbidelli\inst{5},
S. Zaggia\inst{15}
}
   \institute{Institute of Theoretical Physics and Astronomy, Vilnius University,
              Saul\.{e}tekio av. 3, LT-10257 Vilnius, Lithuania\\
              \email{grazina.tautvaisiene@tfai.vu.lt } 
     \and
INAF – Osservatorio di Astrofisica e Scienza dello Spazio, via Gobetti 93/3, I-40129 Bologna, Italy  
\and
Department of Astrophysics, School of Physics, University of New South Wales, Sydney, NSW 2052, Australia 
\and
UNSW Data Science Hub, University of New South Wales, Sydney, NSW 2052, Australia 
         \and
         INAF - Osservatorio Astrofisico di Arcetri - Largo Enrico Fermi 5, I-50125 Firenze, Italy 
         \and
        Space Science Data Center, Via del Politecnico SNC, I-00133 Roma, Italy 
         \and       
        Dipartimento di Fisica e Astronomia, Universit\`a degli Studi di Bologna, Via Gobetti 93/2, I-40129 Bologna, Italy 
        \and
         Nicolaus Copernicus Astronomical Center, Polish Academy of Sciences, ul. Bartycka 18, 00-716, Warsaw, Poland 
        \and
         Institute of Astronomy, University of Cambridge, Madingley Road, Cambridge, CB3 0HA, United Kingdom 
         \and
         Lund Observatory, Department of Astronomy and Theoretical Physics, Box 43, SE-221 00 Lund, Sweden 
         \and
         Max-Planck Institut f\"{u}r Astronomie, K\"{o}nigstuhl 17, 69117 Heidelberg, Germany 
         \and
         Leibniz-Institut f\"{u}r Astrophysik Potsdam, An der Sternwarte 16, 14482 Potsdam, Germany 
        \and
        Dipartimento di Fisica e Astronomia, Universita' di Padova, Vicolo Osservatorio 3, I-35122, Padova, Italy 
        \and
         Observational Astrophysics, Department of Physics and Astronomy, Uppsala University, Box 516, 75120 Uppsala, Sweden 
        \and
        INAF - Osservatorio Astronomico di Padova vicolo dell'Osservatorio, 5, I-35122, Padova, Italy 
             }

   \date{Received 17 September 2021/ Accepted 10 November 2021}

\authorrunning{G. Tautvai\v{s}ien\.{e} et al.}
\titlerunning {Detailed abundances of red giants in the peculiar globular cluster NGC\,1851}

 
  \abstract
   {NGC\,1851 is one of several globular clusters for which multiple stellar populations of the subgiant branch 
have been clearly identified and a difference in metallicity detected. A crucial piece of information on the formation history of this cluster can be provided by the sum of $A$(C+N+O) abundances. However, these values have lacked a general consensus thus far.  The separation of the subgiant branch can be  based on age and/or $A$(C+N+O) abundance differences.   
}
   {Our main aim was to determine carbon, nitrogen, and oxygen abundances for evolved giants in the globular cluster  NGC\,1851 in order to check whether or not the double populations of stars are coeval. 
}
   {High-resolution spectra, observed with the FLAMES-UVES spectrograph on the ESO VLT telescope, were analysed using a differential 
   model atmosphere method. Abundances of carbon were  derived using spectral synthesis of the ${\rm C}_2$ band heads at
5135 and 5635.5~{\AA}. The wavelength interval 6470--6490~{\AA}, with 
CN features, was analysed to determine nitrogen abundances.  Oxygen abundances were
determined from the [O\,{\sc i}] line at 6300~{\AA}. Abundances of other chemical elements were determined from equivalent widths or spectral syntheses of unblended spectral lines. 
} 
   {We provide abundances of up to 29 chemical elements for a sample of 45 giants in NGC\,1851.
   The investigated stars can be separated into two populations with a difference of 0.07~dex in the mean metallicity, 0.3~dex in the mean C/N, and 0.35~dex in the mean $s$-process dominated element-to-iron abundance ratios [$s$/Fe]. No significant difference was determined in the mean values of $A$(C+N+O) as well as in abundance to iron ratios of carbon, $\alpha$- and iron-peak-elements, and of europium.   }
  {As the  averaged $A$(C+N+O) values between the two populations do not differ, additional  evidence is given that NGC\,1851 is composed of two clusters, the metal-rich cluster being by about 0.6~Gyr older than the metal-poor one. A global overview of NGC\,1851 properties and the detailed abundances of chemical elements favour its formation in a dwarf spheroidal galaxy that was accreted by the Milky Way. }


   \maketitle

\section{Introduction}

Chemical abundances are an excellent tool for studying the star-formation and self-enrichment histories of Galactic globular clusters. They trace the integrated nucleosynthetic history of the material in the stars: the early supernovae that raised their overall abundance levels into the typical range for globular clusters, the stellar-mode feedback that imprinted the light-element anticorrelations found in every Galactic globular cluster, and the internal processing that marks specific stages of stellar evolution.

Different elements participate in these processes through different nucleosynthetic channels. The iron-peak and $\alpha$-elements are produced in the early supernovae. The heavy neutron-capture elements are produced in supernovae, in asymptotic giant branch stars, and during mergers of neutron stars and black holes. The light elements from lithium to silicon are produced and destroyed in proton-capture reactions at temperatures typically found in stellar interiors, meaning that their abundance in a star will be a result of stellar feedback into the interstellar medium before it formed as well as the first dredge-up and deep mixing during its lifetime.

The CNO, NeNa, and MgAl cycles, operating in equilibrium, produce anticorrelations in elemental abundances because of the different timescales of the different proton-capture reactions. Carbon, oxygen, and magnesium are depleted, while nitrogen, sodium, and aluminium are produced. This is the same pattern that is observed within globular clusters, which is why it is presumed that globular cluster abundance anomalies could be a result of chemical feedback between multiple generations of star formation early in the history of globular clusters (see reviews by  \citealt{Bastian18} and \citealt{Gratton19}). 

An intriguing field of research was opened up thanks to the  discovery that some globular clusters have stellar populations that differ in metallicity and abundances of slow neutron-capture process  ($s$-process) elements (e.g. \citealt{Carretta10}; \citealt{Marino11}; \citealt{Lardo13}; \citealt{Mucciarelli15}; \citealt{Lee16}; \citealt{Kovalev19}; and references therein). These discoveries imply that those clusters must have had a more complex formation history than ``normal'' globular clusters and may even have originated as nuclear star clusters within dwarf galaxies that were captured by the Milky Way in the past (e.g. \citealt{Bekki12, Massari19}).

Since the CNO, NeNa, and MgAl cycles operate at successively higher temperatures, they are not always well-coupled in a particular cluster. The CNO cycle is the most universal, since it requires the lowest temperature, while the MgAl cycle and the production of Si through the reaction of $^{\rm 27}{\rm Al(p,\gamma)}^{\rm 28}{\rm Si}$ (see \citealt{Yong05} and \citealt{Ventura11} for observational and theoretical discussion of this phenomenon) is only sporadically found in globular clusters. As an example, the range of [Mg/Fe] variation in NGC\,6752 is quite small, even though the O-Na anti-correlation is clear \citep{Yong08}. \citet{Pancino17a} demonstrated that the extent of the Mg-Al anticorrelation depends on both the mass and metallicity of globular clusters.
However, the entire range of anticorrelation behaviour is not well understood because the full abundance pattern from carbon through aluminium is very rarely reported in a single study. This might be due to a technicality: O, Na, Mg, and Al abundances are always derived from resonance lines in high-resolution spectra, while C and N abundances are often derived from molecular features in low-resolution spectra.

Interpretations of carbon and nitrogen abundances can be further complicated by the fact that they can change during a star's lifetime due to internal processes. The first dredge-up is a short-lived phase that occurs shortly after a star has ended core hydrogen burning and begun to transit onto the red giant branch (RGB). During the first dredge-up, a star's convective envelope temporarily expands, spreading into regions that have undergone nuclear burning and then retreating outward. Because the mixing timescale of the convective envelope is quite short, this episode homogenises all material within the envelope, causing the surface abundances to show signs of having undergone hot hydrogen burning. Specifically, the first dredge-up causes a noticeable decrease in [C/Fe], a comprehensive increase in [N/Fe], and a dramatic reduction in the abundance of lithium and the isotopic ratio $^{\rm 12}{\rm C/}^{\rm 13}{\rm C}$. 

Deep mixing is a non convective circulation process that begins when an RGB star's hydrogen-burning shell crosses the deepest point reached by the first dredge-up and continues for the entire RGB lifetime of the star. The exact mechanism for deep mixing is not known, and theoretical arguments have been made for: rotational mixing (\citealt{Sweigart79, Palacios03, Chaname05, Denissenkov06}), magnetic fields (\citealt{Hubbard80, Busso07, Nordhaus08, Palmerini09}), rotation and magnetic fields (\citealt{Eggengerger05}), internal gravity waves (\citealt{Zahn97, Denissenkov00}), thermohaline mixing (\citealt{Eggleton06, Eggleton08, Charbonnel07, Cantiello10, Charbonnel10}), combination of thermohaline mixing and magnetic fields \citep{Busso07, Denissenkov09}, and a combination of thermohaline mixing and rotation (\citealt{Charbonnel10, Lagarde12}). Thus, when investigating and comparing CNO abundances in evolved stars of globular clusters, the stars  ought to be of the same evolutionary stage.  

It has been suspected long ago that NGC\,1851 is chemically heterogeneous (\citealt{Hesser82, Grundahl06}). The high-precision photometric study with the Hubble Space Telescope by \citet{Milone08} revealed two distinct subgiant branches in NGC\,1851 and made this cluster  of particular interest. However, explanations of their origin lack a consensus thus far. 
Ancient globular clusters are usually thought to host two generations of stars. The first is likely primordial and the second born from the ejecta of a fraction of the stars of the first population. Thus, it was initially thought that this could also be the origin of the two subgiant branches in NGC\,1851.
(e.g. \citealt{Pancino10}). Alternatively, there were suggestions that NGC\,1851 originated via the merging of two globular clusters (e.g. \citealt{Campbell12})
or is a naked nucleus of a captured and disrupted dwarf galaxy (e.g. \citealt{Bekki12}; 
\citealt{Marino14}). At first, \citet{vandenBergh96}, and later \citet{Carretta10}, joined the two hypotheses into one and suggested that NGC\,1851 may have been formed by the merger between parental globulars that were once located within a dwarf spheroidal galaxy. 

A crucial piece of information about the NGC\,1851 stellar subcomponents can be brought by way of a robust investigation of $A$(C+N+O) abundances. The importance of $A$(C+N+O) abundances was pointed out already by \citet{Milone08} and by many other investigators (e.g. \citealt{Cassisi08, Villanova10, Alves-Brito12, Gratton12, Yong15, Meszaros21} and references therein). With knowledge of the $A$(C+N+O) abundance, it is possible to answer the question whether a spread in the subgiant branch is caused by the difference in $A$(C+N+O) or in age, as He has a small effect and the required difference in [Fe/H] would show up elsewhere in the colour–magnitude diagram (c.f. \citealt{Gratton12rev}).    

In this study, we investigate abundances of CNO and 26 other chemical elements for a sample of 45 RGB stars of NGC\,1851 
in a framework of the Gaia-ESO Public Spectroscopic Survey (\citealt{Gilmore12, Randich13}). The detailed elemental content of  NGC\,1851 stars is a crucial observational constraint to test the proposed globular cluster formation scenarios. 

The structure of the paper is as follows. In Section~2, we present the observational data and analysis
methods. In Section~3 we discuss the results of the abundance analysis. Our conclusions are drawn in Section~4.

  
\begin{figure}
\includegraphics[width=0.50\textwidth]{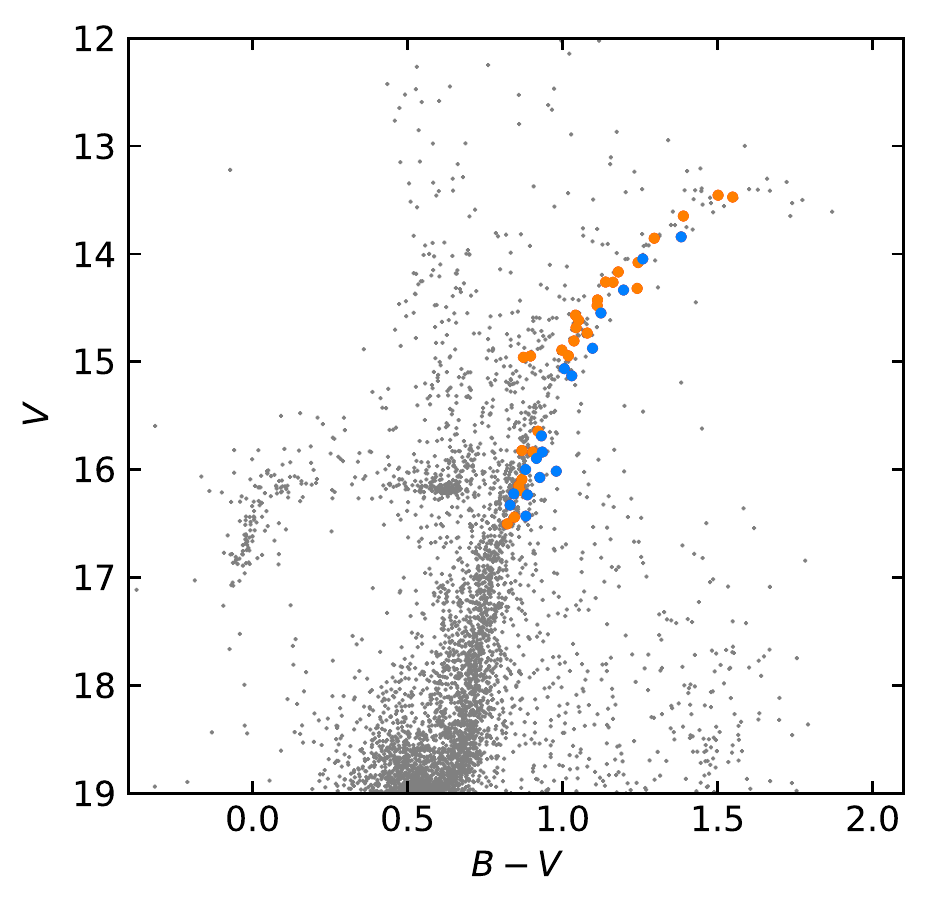}
\caption{$V, B-V$ colour-magnitude diagram of NGC\,1851 (grey circles). Superimposed as larger blue and orange filled circles are the giants investigated in this study. The blue symbols are for the metal-rich subsample stars and the orange symbols are for the metal-poor stars (see the text for the description of how the stars were separated into these subsamples). 
The photometric data were taken from \citet{Stetson19}; namely, the earlier database version corresponding to the one presented in \citealt{Monelli13}.} 
\label{Fig1}
\end{figure}

\begin{table*}
\caption{Main data on the NGC\,1851 cluster target stars}
\label{Table:1}
\centering
\begin{tabular}{cccccccrrc}
\hline\hline
\noalign{\smallskip}
GES ID  &       ID*     &       RA      &       Dec     &       \textit{V}      &       \textit{B}      &       RV      &       S/N     &       S/N     &       \textit{v} sin \textit{i}  \\
        &               &       (J2000) &       (J2000) &       mag     &       mag     &       km s$^{-1}$        &       blue**  &       red**   &       km s$^{-1}$     \\
\hline
\noalign{\smallskip}
\multicolumn{10}{c}{Metal-poor} \\
\noalign{\smallskip}
05133868-4007395        &       14080   &       78.4112 &       $-40.1276$      &       14.320  &       15.560  &       319.92$\pm0.57$ &       58      &       78      &       4.3$\pm0.6$     \\
05134936-4003160        &                       &       78.4557 &       $-40.0544$      &       13.473  &       15.021  &       321.47$\pm0.57$ &       99      &       161     &       6.5$\pm0.9$     \\
05135303-4000431        &       43466   &       78.4710 &       $-40.0120$      &       14.080  &       15.323  &       320.65$\pm0.57$ &       122     &       213     &       4.2$\pm0.3$     \\
05135346-3958452        &       46958   &       78.4728 &       $-39.9792$      &       16.193  &       17.054  &       316.73$\pm0.12$ &       31      &       43      &       3.8$\pm0.5$     \\
05135765-4004312        &       25037   &       78.4902 &       $-40.0753$      &       16.091  &       16.959  &       323.39$\pm0.11$ &       44      &       63      &       3.8$\pm0.5$     \\
05135868-4000122        &                       &       78.4945 &       $-40.0034$      &       14.945  &       15.963  &       325.03$\pm0.57$ &       75      &       105     &       3.9$\pm0.5$     \\
05135918-4002496        &                       &       78.4966 &       $-40.0471$      &       14.959  &       15.832  &       311.14$\pm0.57$ &       95      &       155     &       6.5$\pm0.8$     \\
05135946-4005226        &                       &       78.4978 &       $-40.0896$      &       14.734  &       15.813  &       315.63$\pm0.57$ &       44      &       63      &       3.6$\pm0.7$     \\
05135977-4002009        &       37759   &       78.4990 &       $-40.0336$      &       14.615  &       15.666  &       313.47$\pm0.57$ &       65      &       82      &       4.2$\pm0.3$     \\
05140044-4004135        &       26271   &       78.5018 &       $-40.0704$      &       13.456  &       14.957  &       326.37$\pm0.57$ &       126     &       241     &       6.2$\pm0.5$     \\
05140069-4003242        &                       &       78.5029 &       $-40.0567$      &       14.568  &       15.609  &       321.23$\pm0.57$ &       57      &       90      &       4.0$\pm0.5$     \\
05140180-4002525        &                       &       78.5075 &       $-40.0479$      &       14.684  &       15.727  &       311.70$\pm0.57$ &       43      &       61      &       4.0$\pm0.2$     \\
05140260-4000223        &                       &       78.5108 &       $-40.0062$      &       15.644  &       16.564  &       320.47$\pm0.11$ &       50      &       72      &       4.0$\pm0.3$     \\
05140778-4001183        &       41198   &       78.5324 &       $-40.0218$      &       14.807  &       15.843  &       317.20$\pm0.57$ &       51      &       79      &       3.7$\pm0.5$     \\
05140850-4005545        &       21453   &       78.5354 &       $-40.0985$      &       15.837  &       16.740  &       316.59$\pm0.11$ &       43      &       59      &       4.3$\pm0.3$     \\
05141054-4003192        &                       &       78.5439 &       $-40.0553$      &       14.261  &       15.399  &       315.90$\pm0.57$ &       59      &       80      &       4.5$\pm0.3$     \\
05141057-4003308        &                       &       78.5440 &       $-40.0586$      &       14.947  &       15.843  &       320.22$\pm0.57$ &       93      &       147     &       5.9$\pm0.5$     \\
05141074-4004189        &       25859   &       78.5448 &       $-40.0719$      &       14.263  &       15.425  &       329.22$\pm0.57$ &       76      &       95      &       4.1$\pm0.2$     \\
05141171-3959545        &       45413   &       78.5488 &       $-39.9985$      &       16.151  &       17.009  &       313.40$\pm0.11$ &       36      &       52      &       4.0$\pm0.3$     \\
05141415-3957357        &       47698   &       78.5590 &       $-39.9599$      &       15.823  &       16.691  &       319.49$\pm0.57$ &       40      &       57      &       4.4$\pm0.3$     \\
05141453-4002135        &       36584   &       78.5605 &       $-40.0371$      &       14.479  &       15.590  &       320.90$\pm0.57$ &       54      &       76      &       4.1$\pm0.3$     \\
05141576-4003299        &                       &       78.5657 &       $-40.0583$      &       14.167  &       15.346  &       320.05$\pm0.57$ &       115     &       194     &       4.0$\pm0.2$     \\
05141957-4004055        &                       &       78.5815 &       $-40.0682$      &       14.427  &       15.539  &       327.83$\pm0.57$ &       72      &       96      &       4.5$\pm0.8$     \\
05141979-4006446        &       20426   &       78.5825 &       $-40.1124$      &       16.441  &       17.285  &       319.97$\pm0.12$ &       30      &       45      &       3.8$\pm0.5$     \\
05141988-4003234        &       30286   &       78.5828 &       $-40.0565$      &       13.855  &       15.150  &       318.66$\pm0.57$ &       96      &       169     &       4.9$\pm0.5$     \\
05142070-4004195        &                       &       78.5863 &       $-40.0721$      &       16.503  &       17.324  &       320.74$\pm0.12$ &       28      &       43      &       3.8$\pm0.3$     \\
05142530-4001361        &       39801   &       78.6054 &       $-40.0267$      &       13.649  &       15.038  &       327.17$\pm0.57$ &       93      &       130     &       5.8$\pm0.5$     \\
05142875-4003159        &                       &       78.6198 &       $-40.0544$      &       14.892  &       15.889  &       325.28$\pm0.57$ &       44      &       62      &       4.1$\pm0.3$     \\
\hline
\noalign{\smallskip}
\multicolumn{10}{c}{Metal-rich} \\
\noalign{\smallskip}
05134382-4001154        &       16120   &       78.4326 &       $-40.0209$      &       15.999  &       16.879  &       317.70$\pm0.11$ &       50      &       65      &       3.9$\pm0.8$     \\
05134740-4004098        &       26552   &       78.4475 &       $-40.0694$      &       16.073  &       16.999  &       320.96$\pm0.11$ &       34      &       50      &       3.9$\pm0.3$     \\
05135030-4002071        &       37220   &       78.4596 &       $-40.0353$      &       14.875  &       15.971  &       321.44$\pm0.57$ &       41      &       60      &       4.0$\pm0.6$     \\
05135599-4004536        &       23765   &       78.4833 &       $-40.0816$      &       15.837  &       16.771  &       319.50$\pm0.10$ &       36      &       53      &       3.6$\pm0.6$     \\
05135634-4003448        &       28520   &       78.4848 &       $-40.0624$      &       14.047  &       15.305  &       326.18$\pm0.57$ &       73      &       117     &       4.6$\pm0.6$     \\
05135884-4003365        &       29203   &       78.4951 &       $-40.0601$      &       15.130  &       16.159  &       317.01$\pm0.10$ &       80      &       137     &       4.2$\pm0.3$     \\
05135900-3959591        &       45257   &       78.4958 &       $-39.9998$      &       16.431  &       17.312  &       321.94$\pm0.57$ &       31      &       45      &       4.0$\pm0.6$     \\
05140376-4001458        &                       &       78.5157 &       $-40.0294$      &       15.064  &       16.069  &       315.93$\pm0.57$ &       80      &       137     &       4.4$\pm0.3$     \\
05141447-4001109        &       41689   &       78.5603 &       $-40.0197$      &       14.335  &       15.531  &       321.87$\pm0.57$ &       64      &       93      &       4.1$\pm0.2$     \\
05141566-4000059        &       45006   &       78.5653 &       $-40.0016$      &       16.234  &       17.120  &       314.78$\pm0.11$ &       33      &       50      &       3.9$\pm0.3$     \\
05141615-4001502        &                       &       78.5673 &       $-40.0306$      &       16.015  &       16.994  &       314.11$\pm0.57$ &       49      &       71      &       3.9$\pm0.3$     \\
05141638-4003542        &       27715   &       78.5683 &       $-40.0651$      &       15.687  &       16.618  &       317.81$\pm0.57$ &       47      &       66      &       3.9$\pm0.3$     \\
05141713-4004039        &       26928   &       78.5714 &       $-40.0678$      &       16.224  &       17.066  &       314.91$\pm0.57$ &       32      &       51      &       4.0$\pm0.6$     \\
05142480-4002227        &       35750   &       78.6033 &       $-40.0396$      &       15.896  &       16.811  &       317.76$\pm0.11$ &       38      &       64      &       3.8$\pm0.3$     \\
05142530-4000583        &                       &       78.6054 &       $-40.0162$      &       16.329  &       17.159  &       319.76$\pm0.57$ &       22      &       36      &       5.8$\pm0.5$     \\
05142597-4002538        &       32903   &       78.6082 &       $-40.0483$      &       13.842  &       15.224  &       321.50$\pm0.57$ &       85      &       126     &       5.2$\pm0.6$     \\
05142892-4004454        &                       &       78.6205 &       $-40.0793$      &       14.548  &       15.671  &       325.23$\pm0.57$ &       53      &       77      &       4.2$\pm0.3$     \\      
\hline                                                                                                                                  
\end{tabular}
\tablefoot{*Star IDs are from Carretta (2011), $V$ and $B$ magnitudes from \citet{Stetson19}. The stars were divided into the metal-poor and metal-rich sub-samples according to criteria described in  Sect.~\ref{3.1}. **The median S/N values per pixel in the blue wavelengths at $\lambda$\,4800--5750~\AA, and in the red wavelenghts at $\lambda$\,5850--6800~\AA, accordingly.}
\end{table*}

\section{Observations and method of analysis}

  
\begin{figure}
\includegraphics[width=0.47\textwidth]{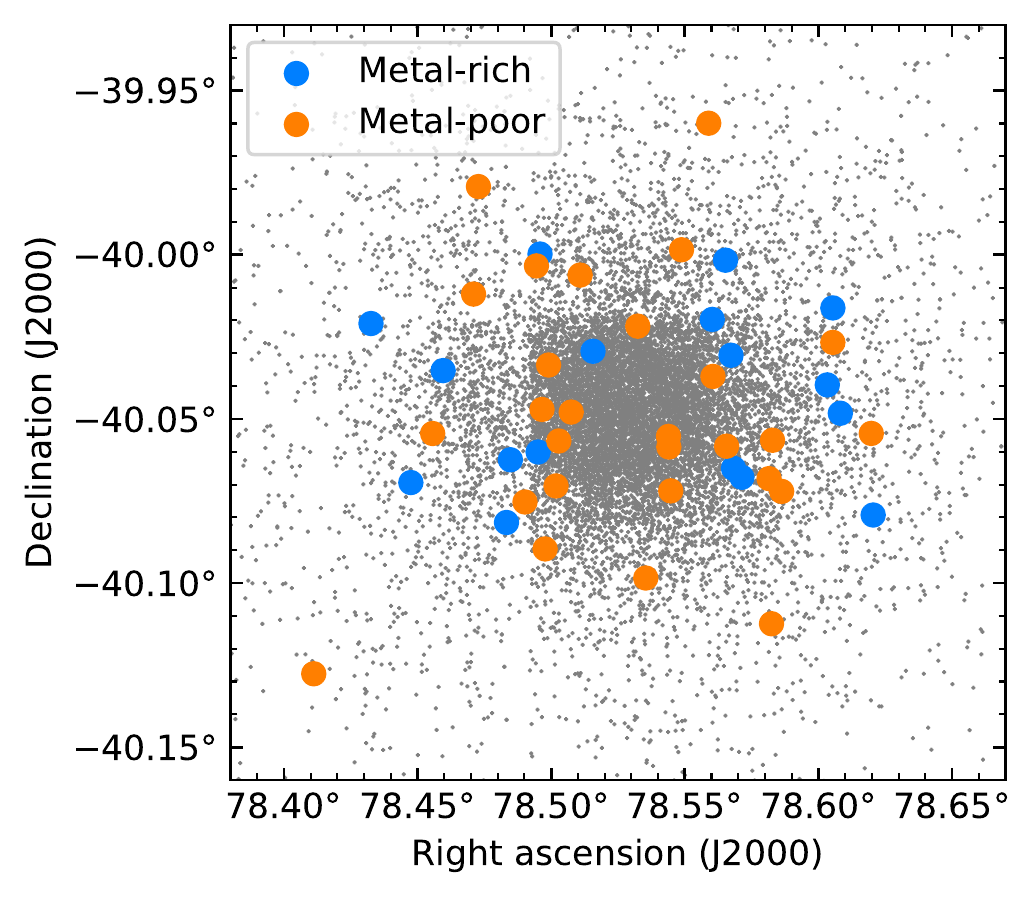}
\caption{Locations of the investigated NGC\,1851 stars (\citealt{Stetson19}). 
The investigated stars are marked by the blue symbols (the metal-rich subsample) and by the orange symbols (the metal-poor stars). } 
\label{Fig2}
\end{figure}

\begin{figure}
 \includegraphics[width=0.49\textwidth]{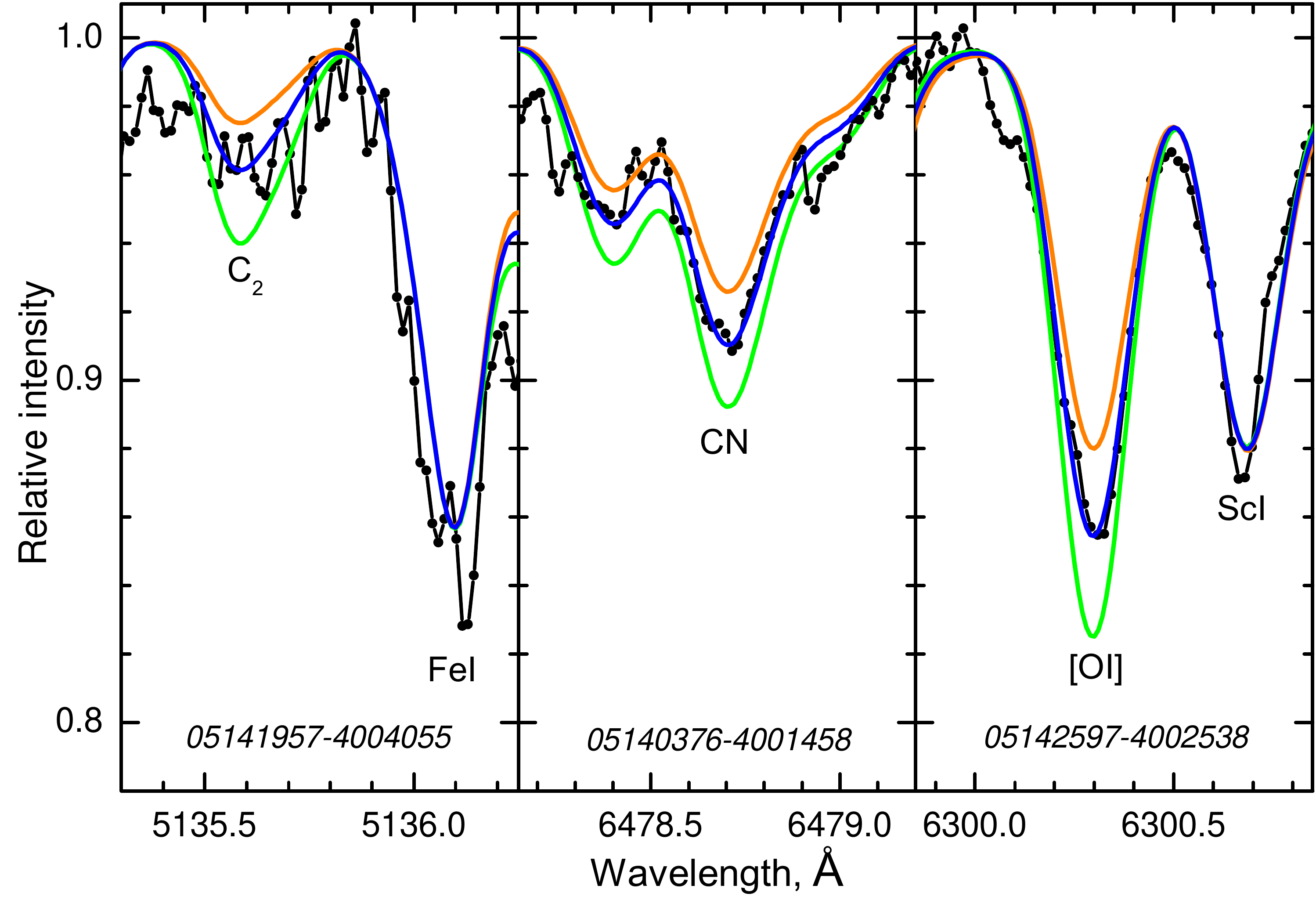}
 \caption{Fits to the 
${\rm C}_2$, 
CN, and forbidden 
[O\,{\sc i}] line in the 
spectra of three NGC\,1851 stars. The observed spectra are shown as black lines with dots. The best-fit synthetic spectra are the blue lines with $\pm 0.1$~dex
changes in corresponding abundances shown as orange and green lines.} 
\label{Fig3}
\end{figure}

 NGC\,1851 is an accessible observational target: relatively close (${\rm D}=12.1$~kpc) and massive ($M=3.2\cdot 10^5M_{\odot}$, \citealt{Baumgardt18}), with low reddening (${\rm E(B-V)}=0.02$, \citealt{Harris96}, 2010 edition). Its halo-like orbit \citep{Dinescu99} results in a radial velocity of $320.5\,{\rm km\,s^{\rm -1}}$, making its stars clearly distinguishable from Galactic field stars in the same line of sight. The target selection strategy is described by \citet{Pancino17b}.  
 
Observations were conducted with the  multi-object instrument  Fibre Large Array Multi-Element Spectrograph (FLAMES, \citealt{Pasquini02}) on the Very Large Telescope (VLT) of the European Southern Observatory (ESO).  
The fibre link to UVES (Ultraviolet and Visual \'{E}chelle Spectrograph, \citealt{Dekker00}) was used to obtain high-resolution spectra ($R = 47\,000$) in a wavelength interval of 4700--6840~{\AA} with a gap of about 50~{\AA} in the centre. For the analysis we used both newly observed and ESO archival spectra from the previous programmes (083.D-0208, 084.D-0470, 084.D-0693, 088.B-0403, 092.D-0171). 
The spectra of the analysed NGC\,1851 stars are stored in the $Gaia$-ESO survey operational database\footnote{The operational database has been developed by the Cambridge Astronomical Survey Unit (CASU) based at the Institute of Astronomy at 
the University of Cambridge. See the website http://casu.ast.cam.ac.uk/gaiaeso/ for more information.}. 

The spectra were reduced with the ESO UVES pipeline and dedicated scripts described by \citet{Sacco14}. Additional efforts were applied in improving their 
continuum normalisation. For this purpose, we used the SPLAT-VO code\footnote{http://star-www.dur.ac.uk/$\sim$pdraper/splat/splat-vo/splat-vo.html}.  
Radial velocities 
 (RV) and rotation velocities ($v\,{\rm sin}\,i$) were also determined by cross-correlating all the spectra with a sample of 
 synthetic templates specifically derived for the Gaia-ESO project. The typical uncertainty on RVs is about 
0.4~km\,s$^{-1}$. A list of the investigated 45 stars in NGC\,1851 is presented in Table~\ref{Table:1}. The stars in 
Table~\ref{Table:1} are divided into the metal-poor and metal-rich sub-samples according to criteria described later in  Sect.~\ref{3.1}. Figure~\ref{Fig1} shows positions of the investigated stars on a colour magnitude diagram and Fig.~\ref{Fig2} their positions around the cluster centre. Information on stellar positions and photometry was taken from a catalogue by \citet{Stetson19}\footnote{https://www.canfar.net/storage/list/STETSON}, the earlier database version  corresponding to the one presented in \citet{Monelli13}.
  
In this study, we use the main stellar atmospheric parameters and abundances of chemical elements presented in the fourth Gaia-ESO survey internal data release. They were determined spectroscopically using a differential model atmosphere technique described by  
\citet{Smiljanic14}. In short, as the Gaia-ESO covers a wide range of targets, the spectra from high-resolution UVES observations were analysed in parallel by several nodes 
of scientists of the Gaia-ESO Consortium. Each of them adopted a slightly different approach or software (or both) with the aim to best cover the whole parameter space. On the other hand, all the nodes used a common line list (\citealt{Ruffoni14, Heiter21}, and references therein), a set of MARCS\footnote{http://marcs.astro.uu.se/} model atmospheres by  \citet{Gustafsson08}, solar abundances by \citet{Grevesse07}, and analysed common calibration targets (\citealt{Pancino17b}).

It was a two-stage approach. First, every node determined the atmospheric parameters. Then, using the Gaia benchmark stars (\citealt{Jofre14}, \citealt{Heiter-Jofre15}), the systematic and random errors affecting each node and correlations among the nodes were inferred by means of a Bayesian MCMC model. This model was then applied to the survey data to estimate the best values of the atmospheric parameters and their uncertainties. After this process, a single set of recommended parameters was provided, and the nodes proceeded to the second step of determining the elemental abundances. In this case, the nodes were asked to provide abundance values for the individual spectral lines of each chemical species that were analysed. The method to homogenise the abundances was similar to the one used for the atmospheric parameters, but there were no reference abundances to rely on. In this case, the repeated measurements in different spectra of the same stars and the measurements of multiple spectral lines (when possible) helped to constrain the correlations and typical uncertainties of each node. A full description of the homogenisation, applied to the last Gaia-ESO data releases, will be described by Worley et al. (in preparation).
Approximate uncertainties of the main atmospheric parameters in the case of NGC\,1851 are $\pm100$~K, $\pm0.25$~dex, and $\pm0.1$~dex for $T_{\rm eff}$, log\,$g$, and [Fe/H], respectively.  

Abundances of carbon and nitrogen were determined using the same method as in \citet{Tautvaisiene15}.  
The ${\rm C}_2$ Swan (1,0) band head at 5135.6~{\AA} and ${\rm C}_2$ Swan (0,1) band head at 
 5635.2~{\AA} were investigated in order to determine the carbon abundance.  
The ${\rm C}_2$ bands are suitable for carbon abundance investigations since they give the same 
carbon abundances as [C\,{\sc i}] 
lines, but are not sensitive to non-local thermodynamic equilibrium (NLTE) deviations (c.f. \citealt{Clegg81, Gustafsson99}). 
Since abundances of C and O are bound together by the molecular equilibrium 
in the stellar atmosphere the oxygen abundance was determined as well. For this purpose, we used the forbidden [O\,{\sc i}] line at 6300.3~\AA. 
Following \citet{Johansson03}, we took into account the oscillator strength values for \textsuperscript{58}Ni and \textsuperscript{60}Ni, which blend the oxygen line. 
Lines of [O\,{\sc i}] are considered as very good indicators of oxygen abundances. It was 
determined that they are 
not only insensitive to NLTE effects, but also give quite similar oxygen abundance results with 3D and 1D model atmospheres (c.f. \citealt{Asplund04, Pereira09}). This line forms in near-LTE and is only weakly sensitive to convection, its formation is similar in 3D radiation hydrodynamic and 3D magneto radition-hydrodynamical solar models (\citealt{Bergemann21}). For the analysis only, we used lines unaffected by the telluric contamination.   
The interval 6470 -- 6490~{\AA} containing $^{12}{\rm C}^{14}{\rm N}$ bands 
was used for the nitrogen abundance analysis. 
All the synthetic spectra have been calibrated to the solar spectrum by \citet{Kurucz05} to make the analysis strictly differential. We used a spectrum synthesis code \textit{Turbospectrum} \citep{Plez12}. 

In fitting the observed and the theoretical spectra, the stellar rotation (provided in Table~\ref{Table:1}) was taken into account.
Approximate values of stellar rotation velocities were evaluated for the Survey stars as described by \citet{Sacco14} and provided in the Gaia-ESO Survey GES iDR4 database. The values of 
$v\,{\rm sin}\,i$ were calculated using an empirical relation of a full width half maximum of the cross-correlation function (CCF$_{\rm FWHM}$) 
and $v\,{\rm sin}\,i$, which was specifically derived for this project. 
Figure~\ref{Fig3} displays examples of spectrum syntheses for the programme stars.

\begin{table}
\centering
\caption{Effects of derived abundances and isotopic ratios for the star NGC\,1851\,05141576-4003299, resulting from
abundance changes of C, N, or O.}
\label{Table:2}
\[
\begin{tabular}{lccc}
\hline
Species & $\Delta$ C & $\Delta$ N & $\Delta$ O \\
& $\pm0.1$~dex & $\pm0.1$~dex &$\pm0.1$~dex \\
\hline
\noalign{\smallskip}
$\Delta$ C                      &       --              &        $\pm0.01$      &        $\pm0.04$       \\
$\Delta$ N                      &       $\mp0.11$       &       --              &        $\pm0.08$       \\
$\Delta$ O                      &       $\pm0.01$       &        $0.00$ &       --              \\
$\Delta$ C/N            &       $\pm0.15$       &        $\mp0.15$      &        0.00    \\
\hline
\end{tabular}
\]
\end{table}

We applied the NLTE corrections to abundances of Li, Na, and Ba, as they have the most sensitive lines. The Li NLTE corrections were computed for every star using the code provided by \cite{Wang21}. The averaged value of all these corrections is $-0.07\pm0.02$~dex. For the sodium abundances, the corrections were taken from \citet{Lind11} and their averaged value is $-0.08\pm0.03$~dex. For the barium abundances, the NLTE corrections were taken from \citet{Korotin14} and their averaged value is $-0.09\pm0.04$~dex.

 We estimated uncertainties on abundances by taking into account the uncertainties in the atomic data, continuum placement variations, and imperfect equivalent width measurements or the fitting of synthetic spectra to each line, as well as uncertainties in the stellar atmospheric parameters. 
 The abundance uncertainties of all the investigated chemical elements for every star are provided along with the abundance results in Table~\ref{Table:3}.   
 
As C, N, and O are linked through molecular equilibrium in the atmospheres of the cool giants we are studying, we must also consider how an uncertainty in one element's abundance would affect the other two. This is summarised in Table~\ref{Table:2} for the star NGC\,1851 05141576-4003299, which is near the median of $T_{\rm eff}$ and log\,$g$ in our sample.
We refer to \citet{Smiljanic14} and \citet{Tautvaisiene15} for more details about the method of analysis and uncertainties.

  
\begin{figure}
\includegraphics[width=0.45\textwidth]{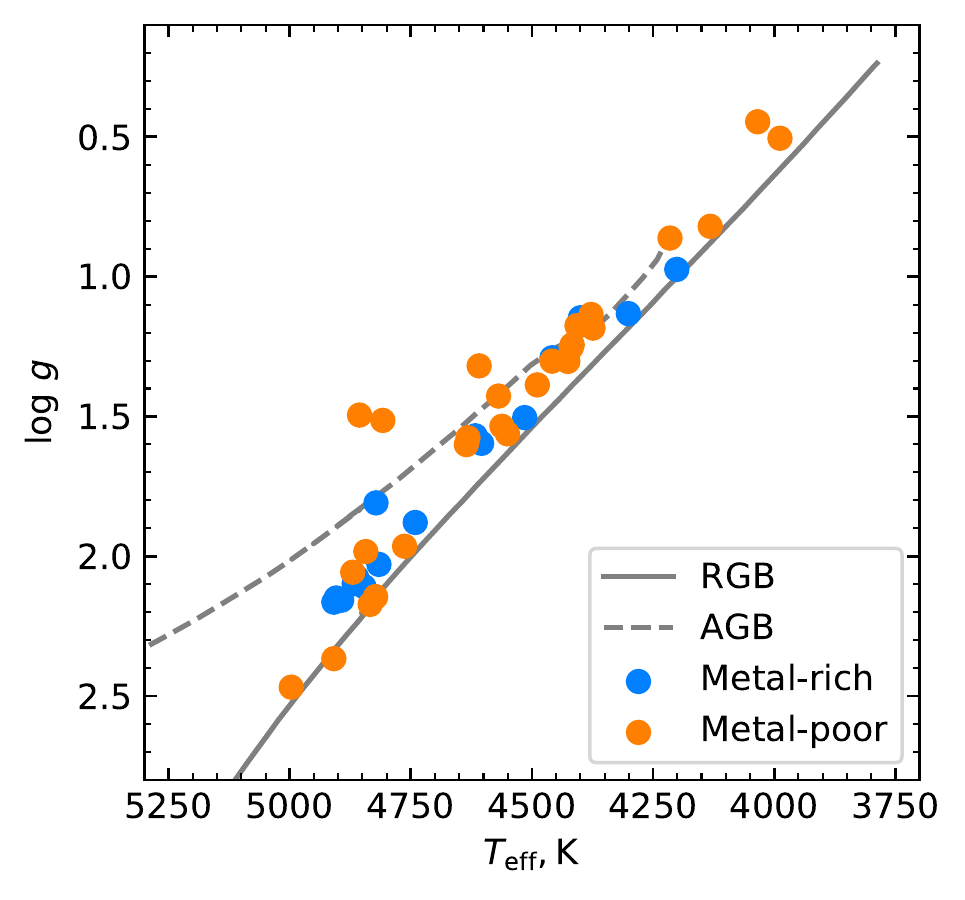}
\caption{Log\,$g$, $T_{\rm eff}$ diagram of the investigated NGC\,1851 stars together with the PARSEC \citep{Bressan12} isochrone of the mean Age = 12.27 Gyr (\citealt{Valcin20}) and [Fe/H] = $-1.0$. 
} 
\label{Fig4}
\end{figure}


\section{Results and discussion}
\label{results}

\begin{figure}
 \includegraphics[width=0.49\textwidth]{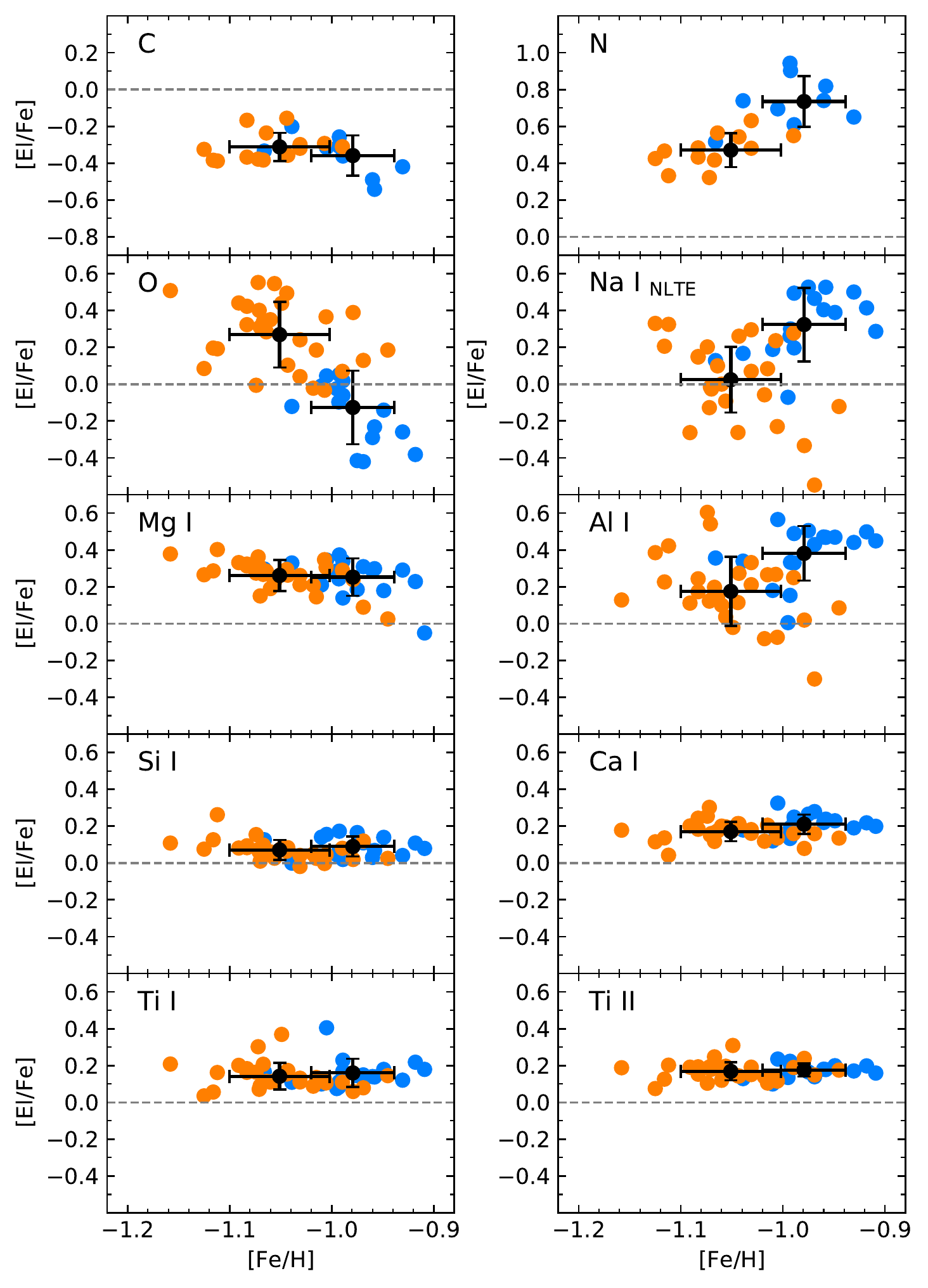}
 \caption{Stellar relative abundances [El/Fe] of light  elements in respect to [Fe/H] for metal-rich (blue symbols) and metal-poor stars (orange symbols) stars. The average values of the metal-rich and metal-poor samples of stars are shown in black. } 
\label{Fig5}
\end{figure}

\begin{figure}
 \includegraphics[width=0.49\textwidth]{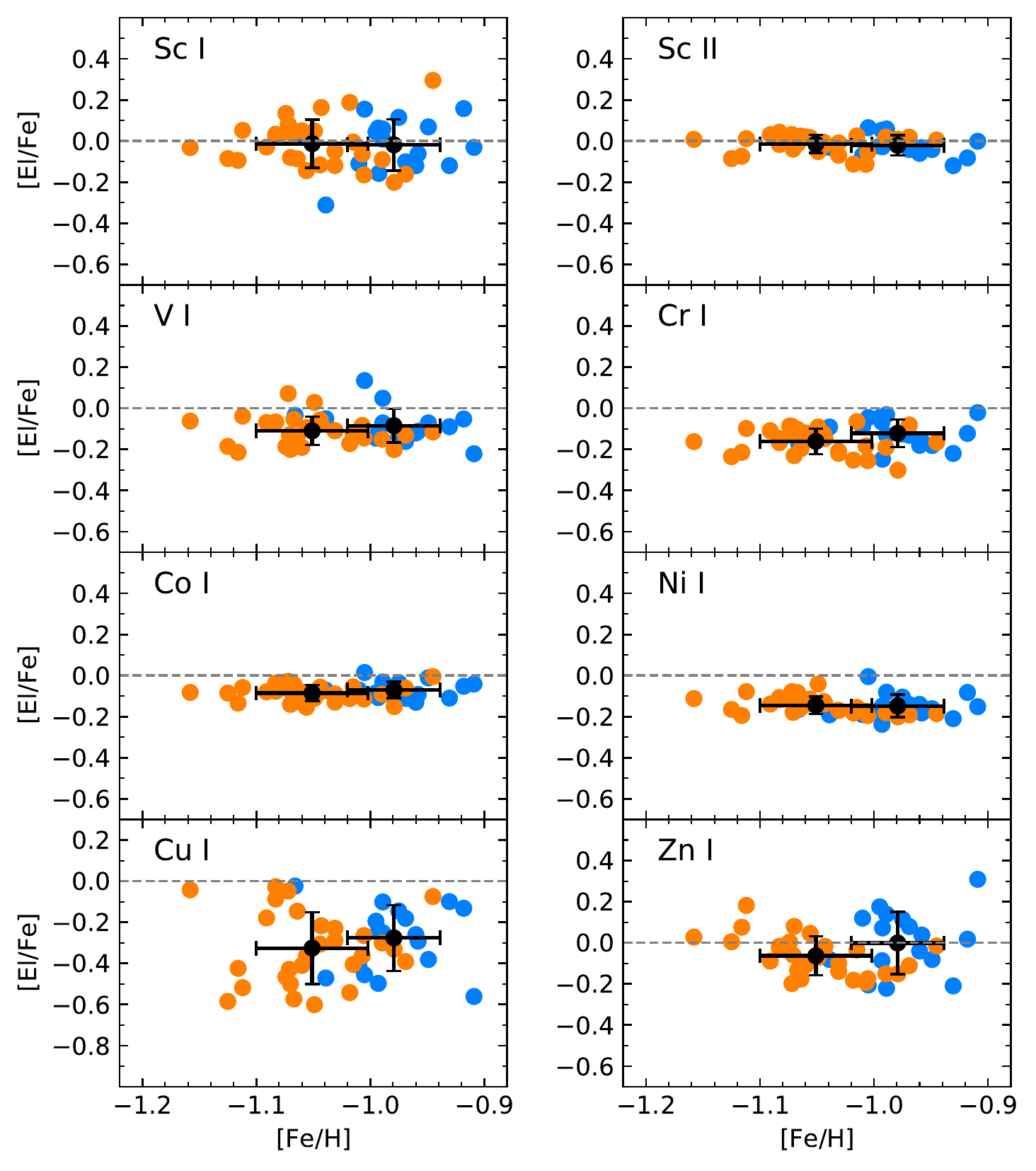}
 \caption{Stellar relative abundances [El/Fe] of iron group elements  in respect to [Fe/H]. } 
\label{Fig6}
\end{figure}

\begin{figure}
 \includegraphics[width=0.49\textwidth]{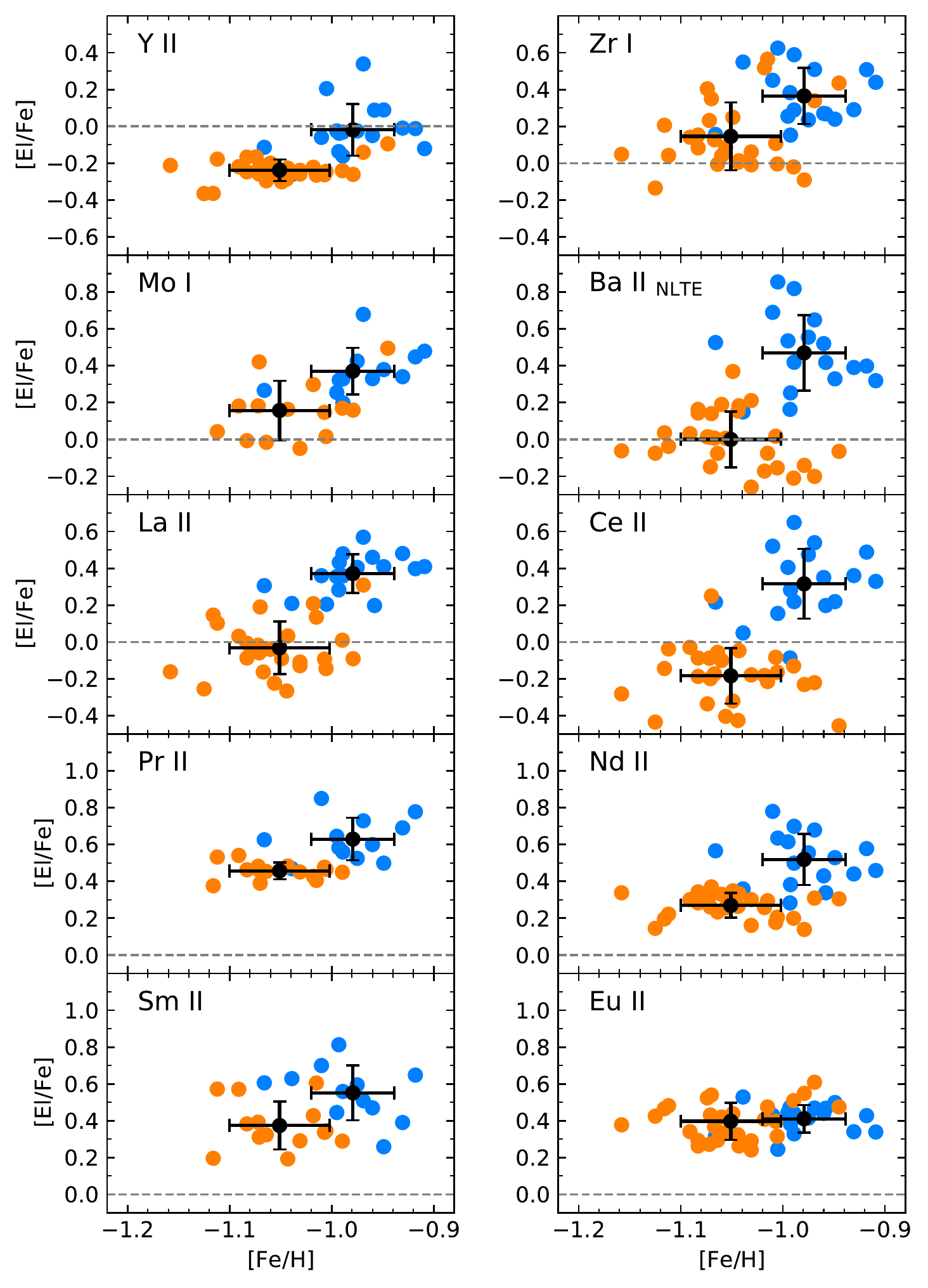}
 \caption{Stellar relative abundances [El/Fe] of neutron-capture elements in respect to [Fe/H]. The meanings of the symbols are the same as in Fig.~\ref{Fig5}.} 
\label{Fig7}
\end{figure}

The main atmospheric parameters and elemental abundances in stars of the globular cluster NGC\,1851 are presented in Table~\ref{Table:3}, a full version of which is provided in an electronic form.
The log\,$g$, $T_{\rm eff}$ diagram of the investigated stars is presented in Fig.~\ref{Fig4} and the determined abundances in Figs.~\ref{Fig5},~\ref{Fig6}, and \ref{Fig7}.

\subsection{Two stellar populations in NGC\,1851}
\label{3.1}
It has been suspected that NGC\,1851 is not chemically homogeneous since the spectroscopic study of \citet{Hesser82}, which found ``extraordinarily strong'' CN bands in three of eight red giant stars, indicating large star-to-star variations probably in nitrogen abundance. Those authors also identified variations in the abundance of Ba and Sr, namely, elements produced by the $s$-process. There have been a number of studies of the complexity in the chemical compositions of stars in NGC\,1851, revealing two populations of stars, very distinct in their nitrogen and $s$-process dominated element abundances.   

\begin{figure}
\centering
 \includegraphics[width=0.40\textwidth]{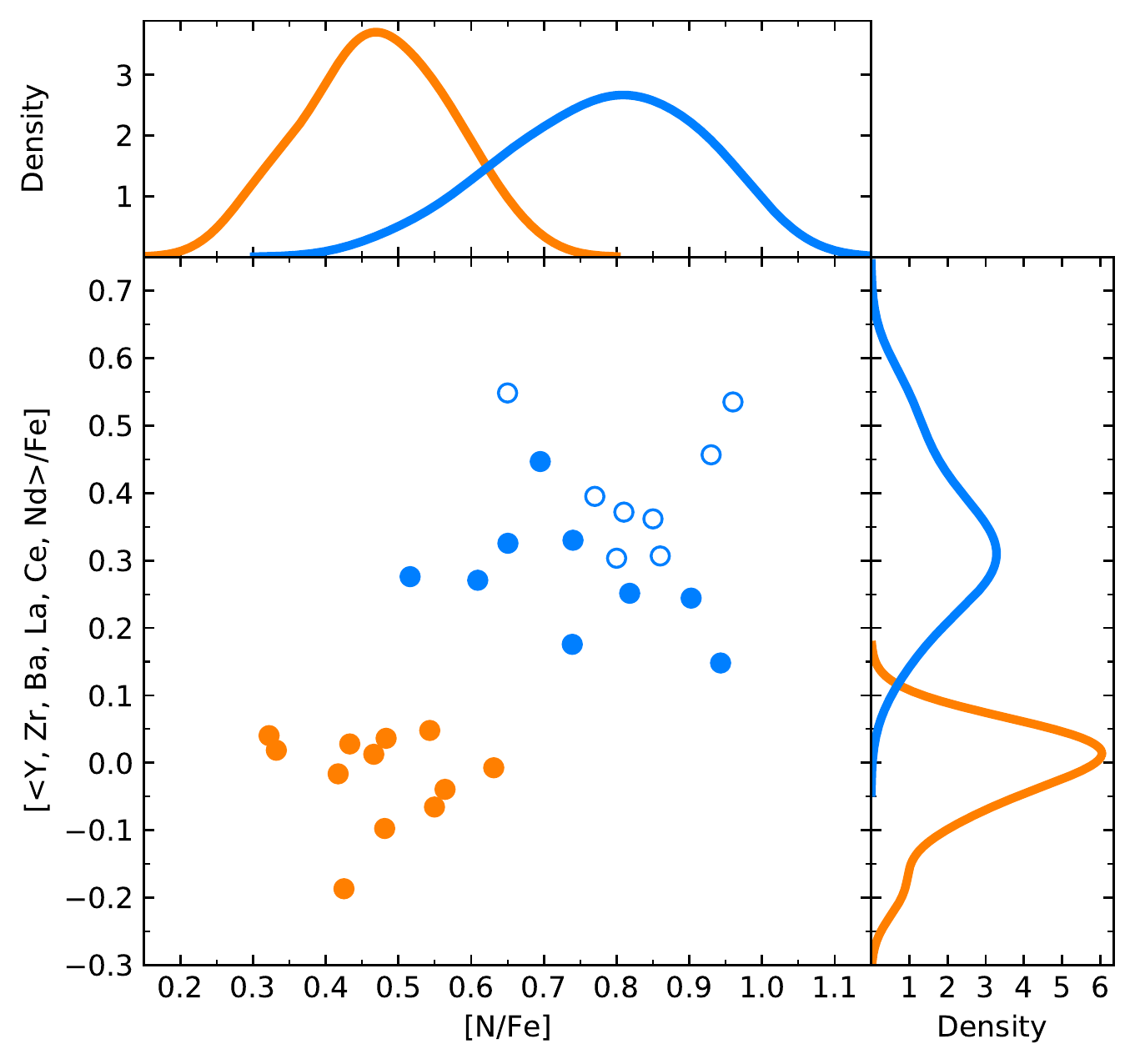}
 \caption{Averaged abundances of the $s$-process dominated chemical elements in respect to [N/Fe] for stars in metal-rich (blue symbols) and metal-poor (orange symbols) populations. The text provides an explanation of the empty blue symbols. } 
\label{Fig8}
\end{figure}

\begin{figure}
\centering
 \includegraphics[width=0.40\textwidth]{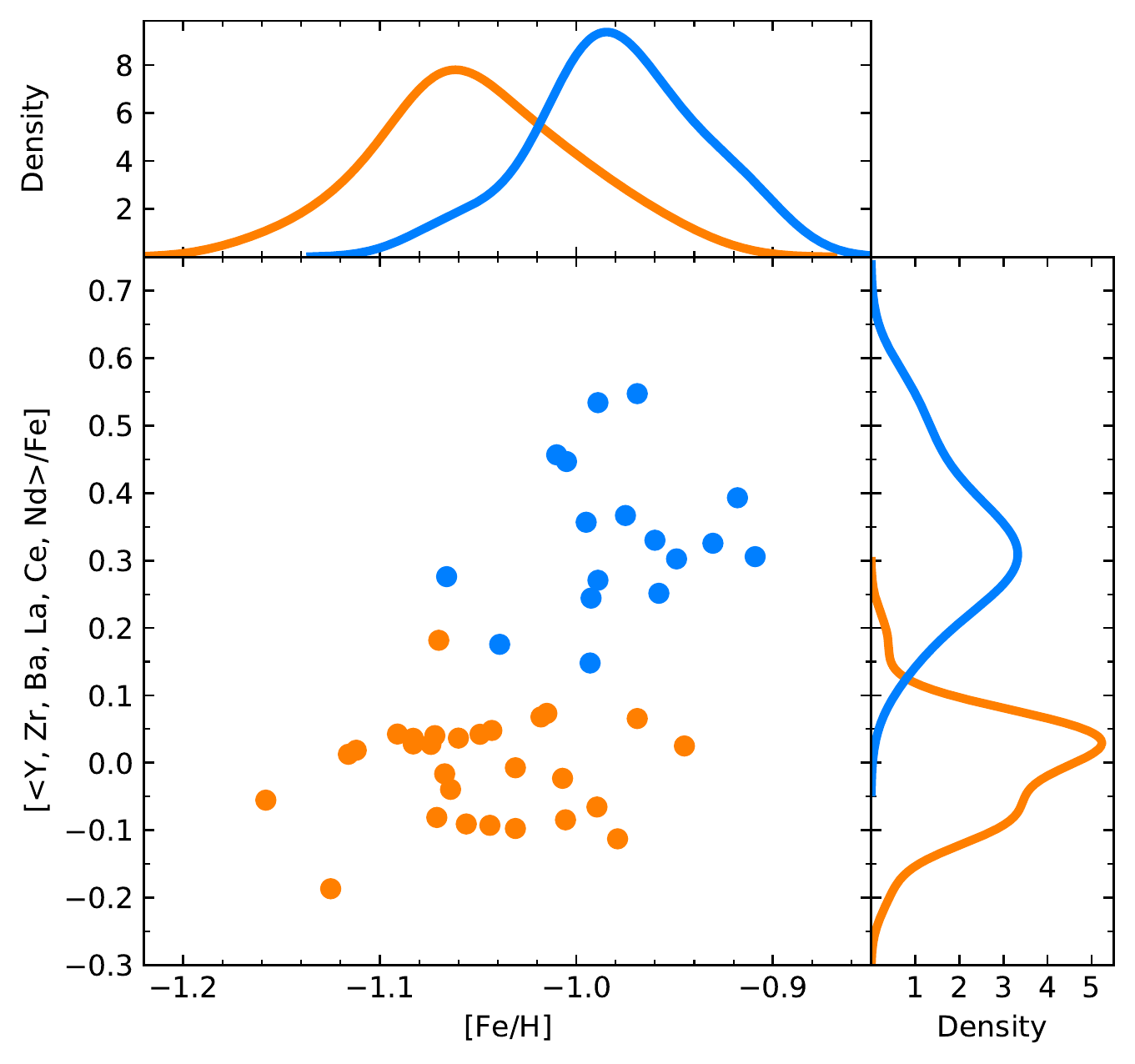}
 \caption{Averaged abundances of the $s$-process dominated chemical elements in respect to [Fe/H]. The meanings of the symbols are the same as in Fig.~\ref{Fig8}.}  
\label{Fig9}
\end{figure}

We separated NGC\,1851 stars between metal-rich and metal-poor ones according to their 
 abundances of iron, nitrogen, and $s$-process dominated elements.  Figure~\ref{Fig8} shows that the stars fall into two clear subsamples according to their mean abundances of $s$-process dominated elements and nitrogen-to-iron ratios. As the nitrogen abundances in GES iDR4 were determined for only part of the sample of stars (since the lines of C$_2$ were too week for the robust carbon abundance determination and, consequently, we were not able to determine the nitrogen abundance from CN lines), we decided to check the strength of CN lines and try to overcome this problem for stars with clearly visible CN lines. We found that nine stars had the CN lines clearly visible. We accepted the mean value of carbon abundance calculated for the subsample of nitrogen rich stars and determined approximate values of their nitrogen abundances. As the scatter of the mean carbon abundance is quite small, those values of nitrogen determination should be rather accurate. We plotted them as empty circles in Fig.~\ref{Fig8}. The stars that had the CN lines almost invisible were attributed to the metal-poor subsample with low nitrogen abundance. The division of the whole sample is shown in Fig.~\ref{Fig9}. The hardest separation between the two subsamples is at the mean value of $s$-process dominated elements of 0.1~dex. Only one star is slightly above this value, however, its nitrogen and iron abundances are low; thus we attributed it to the metal-poor subsample. The stellar density distributions are also presented and we can see the two clear populations of stars. In [N/Fe] and [Fe/H], there is some overlap in abundance distribution tails due to a natural spread and uncertainties.
  In Table~\ref{Table:4}, we present the averaged elemental abundances and their ratios for the two populations of investigated NGC\,1851 stars. Abundance ratios to iron of $\alpha$-elements, iron-peak elements, and the $r$-process-dominated element europium in both subsamples are very similar. However, differences in other chemical elements are noticeable.    
 
  \subsection{Metallicity}
 
 Thus, the investigated stars were divided into the subsamples of 17 metal-rich and 28 metal-poor stars with  averaged metallicities of $-0.98\pm0.04$~dex and $-1.05\pm0.05$~dex, respectively. We performed a significance test for those subsamples and we obtained a $p$-value as low as $1\cdot10^{-5}$, which demonstrates the probability that this [Fe/H] difference could have occurred just by  chance. 
 \citet{Carretta10,Carretta11} have demonstrated that the s-rich and s-poor subsamples of stars in NGC\,1851 also differ in metallicity by about 0.07--0.08~dex. Those authors suggested that NGC\,1851 might be the result of a merger of two globular clusters, as proposed by \citet{vandenBergh96}, and their age difference could be 1~Gyr. A metallicity difference of about 0.07~dex was also determined by \citet{Gratton12}, who analysed giants and horizontal-branch stars.  The mean [Fe/H] values in the metal-rich and metal-poor populations investigated in our work also differ by 0.07~dex, although the absolute metallicity values are slightly different in various studies due to differences in the methods of analysis.  It has been investigated whether the NLTE effects on Fe\,{\sc i} lines could cause the metallicity spread or bimodal iron abundances in some globular clusters (e.g. \citealt{Mucciarelli15, Lee16, Kovalev19}). As we see in Fig.~3 by \citet{Kovalev19}, the NLTE effects in the NGC\,1851 parameters are very small. They can be larger in more metal-deficient globular clusters when stronger spectral lines are used for the analysis.

\subsection{Elements produced by $s$-process}

 In our study, the mean values for two subsamples of stars in averaged $s$-process element abundances are $0.01\pm0.08$~dex and $0.34\pm0.11$~dex for the metal-poor and metal-rich subsamples, respectively. 
\citet{Yong08} confirmed the $s$-process variations noted by \citet{Hesser82} and suggested that the s-rich and s-poor abundance groups might correspond to the two photometrically identified subgiant branches. 
\citet{Gratton12} investigated elemental abundances of stars in the two subgiant branches of NGC\,1851 and found that the fainter one is composed by metal-rich and $s$-process-element-rich stars, and the brighter on has metal-poor and $s$-process-element-poor stars. In our study, as well as that of \citet{Carretta11}, we see the same relation between metallicity and $s$-process dominated element abundances.

\begin{figure}
\centering
 \includegraphics[width=0.47\textwidth]{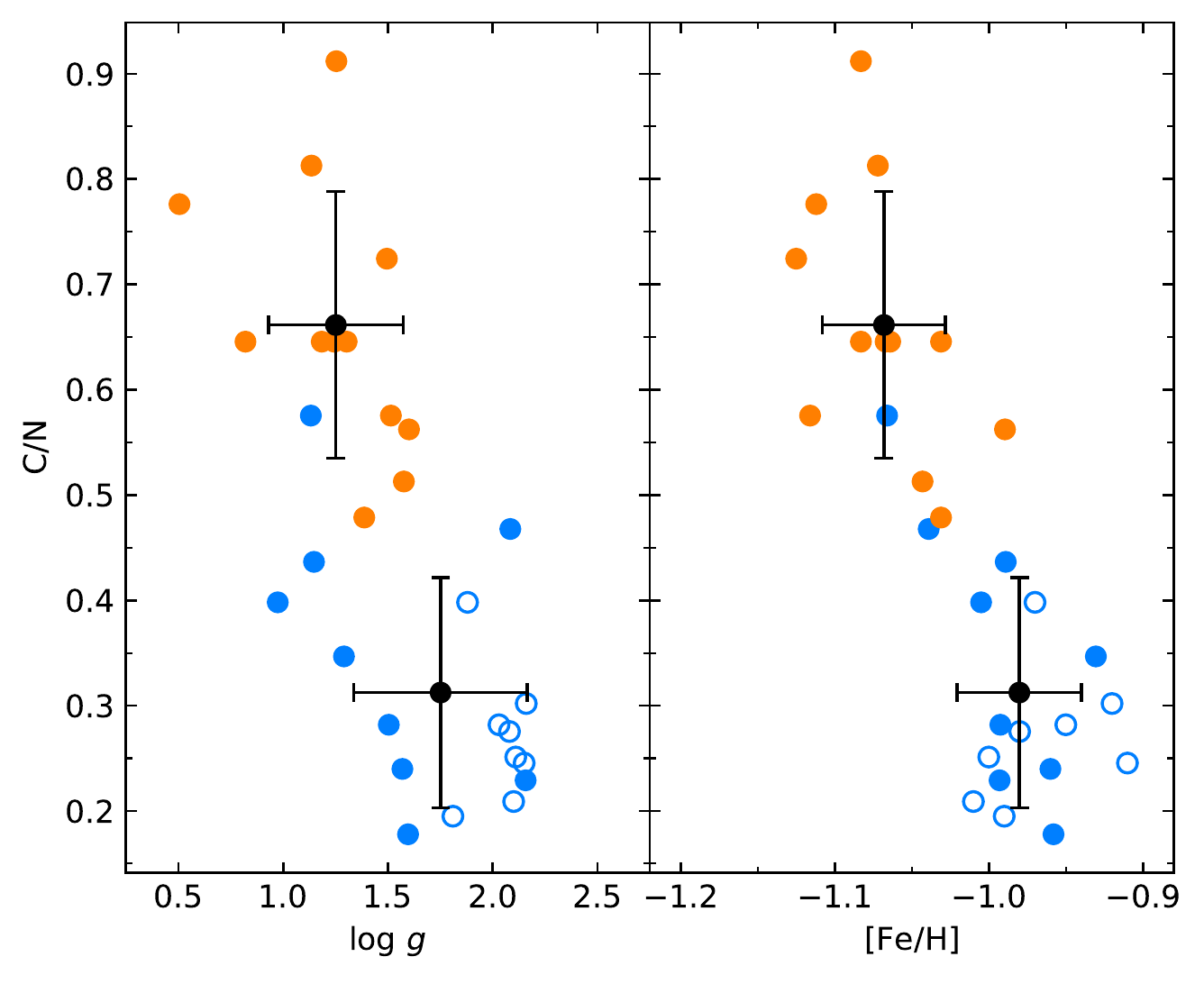}
 \caption{C/N ratios in respect of the surface gravity and [Fe/H] values. The meanings of symbols are the same as in Fig.~\ref{Fig8}.}  
\label{Fig10}
\end{figure}

\begin{figure}
\centering
 \includegraphics[width=0.40\textwidth]{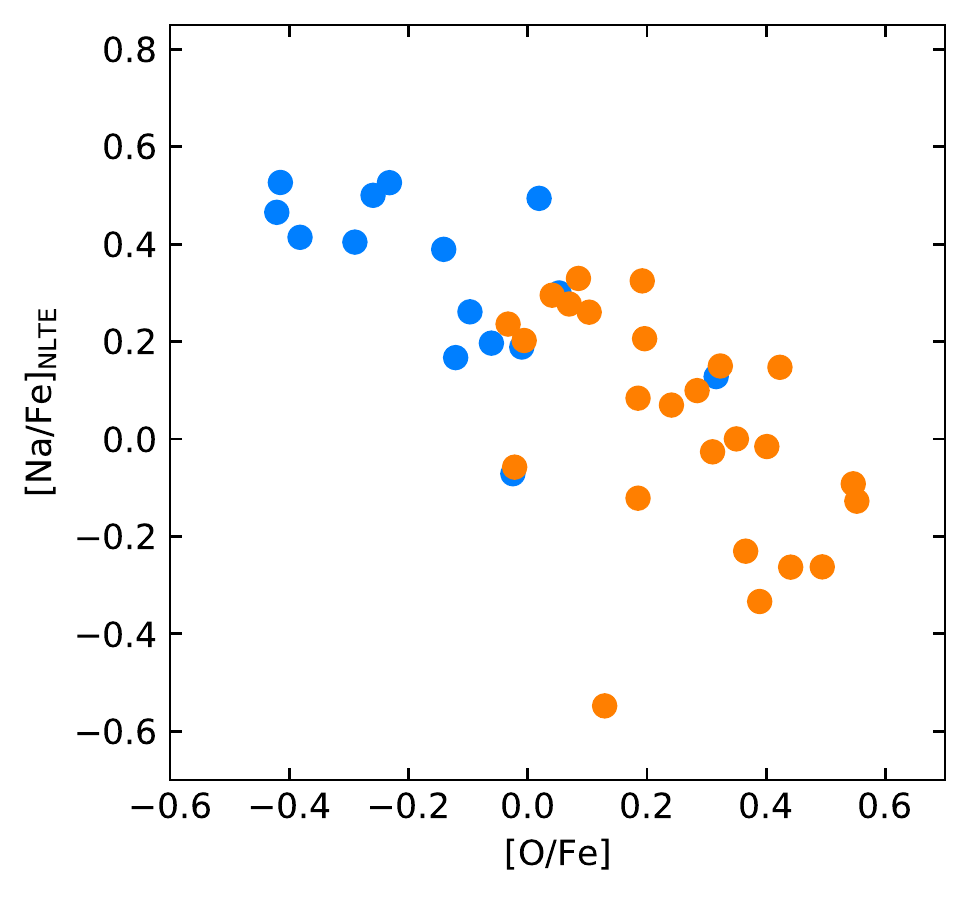}
 \caption{[Na/Fe] in respect to [O/Fe]. The meanings of symbols are the same as in Fig.~\ref{Fig8}.}  
\label{Fig11}
\end{figure}

\subsection{Light elements}

The metal-rich and metal-poor subsamples have very similar carbon abundances but distinct abundances of nitrogen and consequently different values of C/N ratios. Figure\,\ref{Fig10} shows where the sampled stars are located in the C/N versus log\,$g$ and [Fe/H] planes. The mean value of the C/N ratio in the metal-rich stars is $0.35\pm0.13$, and  $0.66\pm0.13$ in metal-poor stars.
In NGC\,1851, \citet{Lim15} also found no difference in strengths of CH lines between the
two groups of stars with different CN strengths, while they found that CN and CH are anticorrelated in NGC\,288 and a positive correlation is in M\,22.
In another Gaia-ESO survey paper, \citet{Lagarde19}  showed that C and N abundances in NGC\,1851 are reproduced well with the thermohaline
mixing only (their Fig.~8).

Differences among the two populations are also present in the abundances of oxygen and sodium  (Fig.~\ref{Fig5}), and both populations exhibit Na-O anticorrelations (Fig.~\ref{Fig11}). 
In stellar samples investigated by \citet{Carretta10,Carretta11}, the Na-O anticorrelation was present to the same degree in both the relatively metal-rich and relatively metal-poor stars.  \citet{Gratton12-ONa} also found a Na-O anti-correlation among blue horizontal branch stars, which partially overlaps that which is found among red horizontal branch stars, although the blue horizontal branch stars were found to be more Na-rich and O-poor overall.

\begin{figure}
 \includegraphics[width=0.47\textwidth]{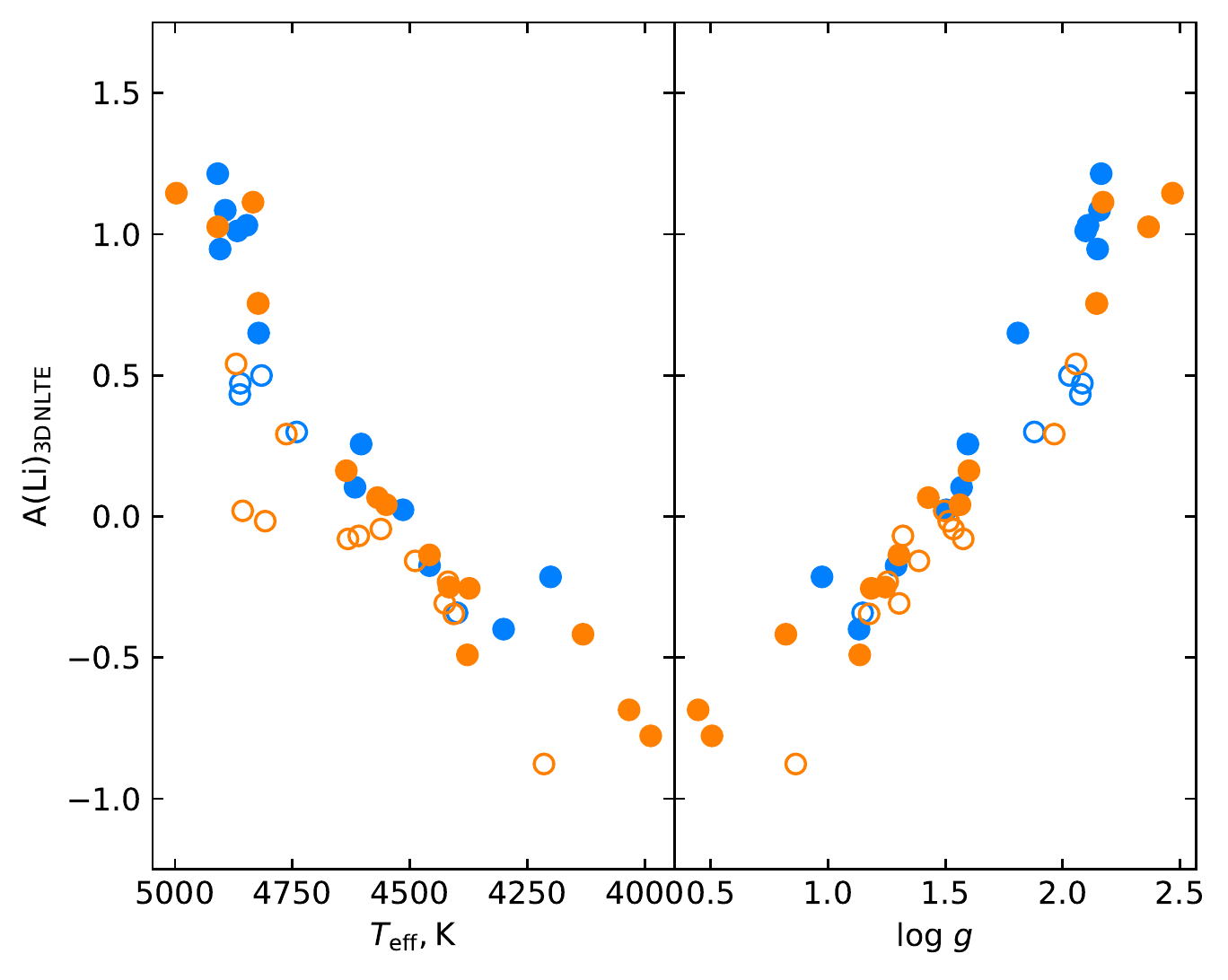}
 \caption{Lithium abundances in respect of the effective temperature and surface gravity values for stars in metal-rich (blue symbols) and metal-poor (orange symbols) populations. The upper values of Li abundances are plotted by empty circles. }
\label{Fig12}
\end{figure}

\begin{figure}
\centering
 \includegraphics[width=0.45\textwidth]{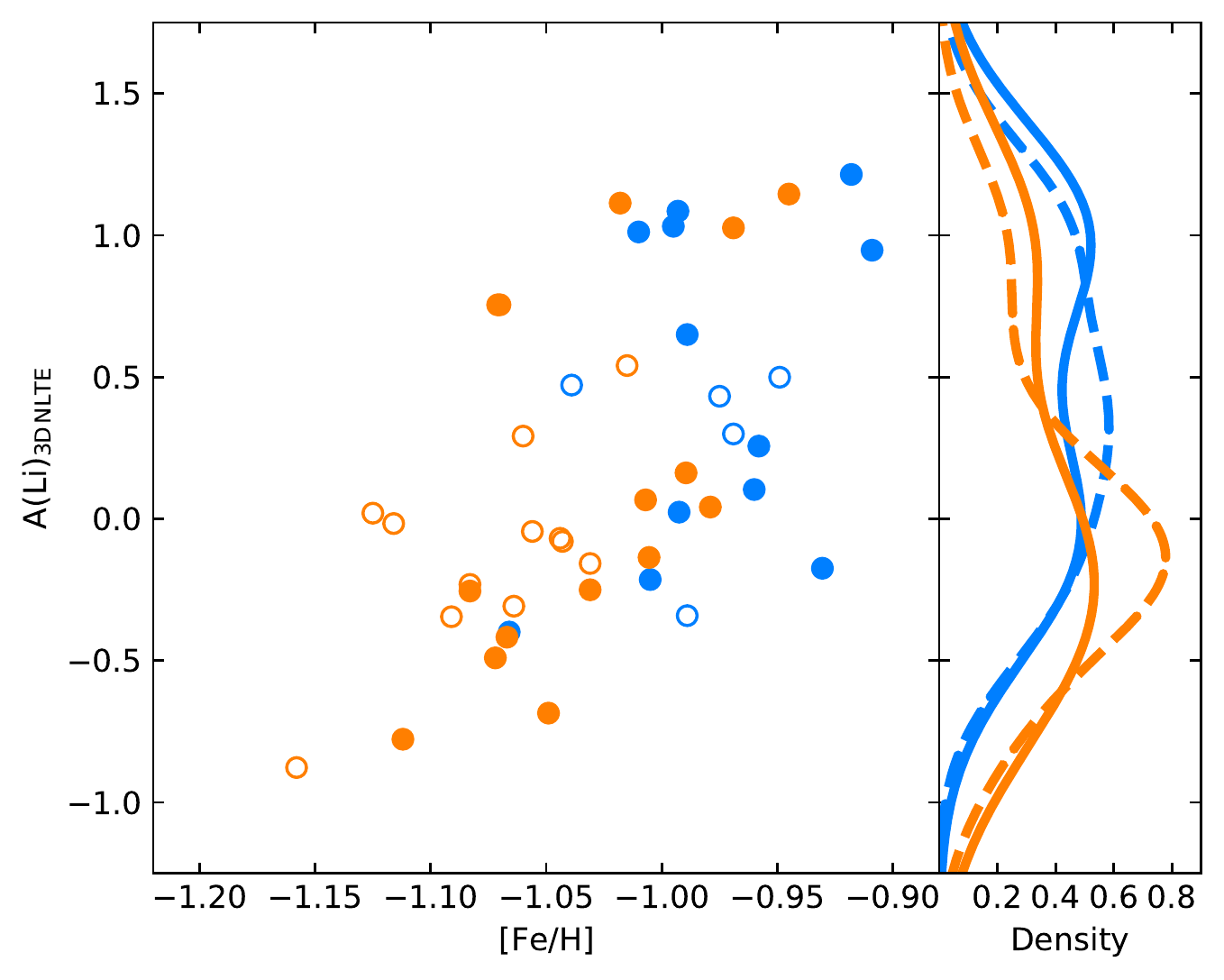}
 \caption{Lithium abundances in respect of metallicity. The meanings of symbols are the same as in Fig.~\ref{Fig12}. The dashed lines indicate densities with the upper Li abundances included.} 
\label{Fig13}
\end{figure}

\begin{figure}
\centering
 \includegraphics[width=0.45\textwidth]{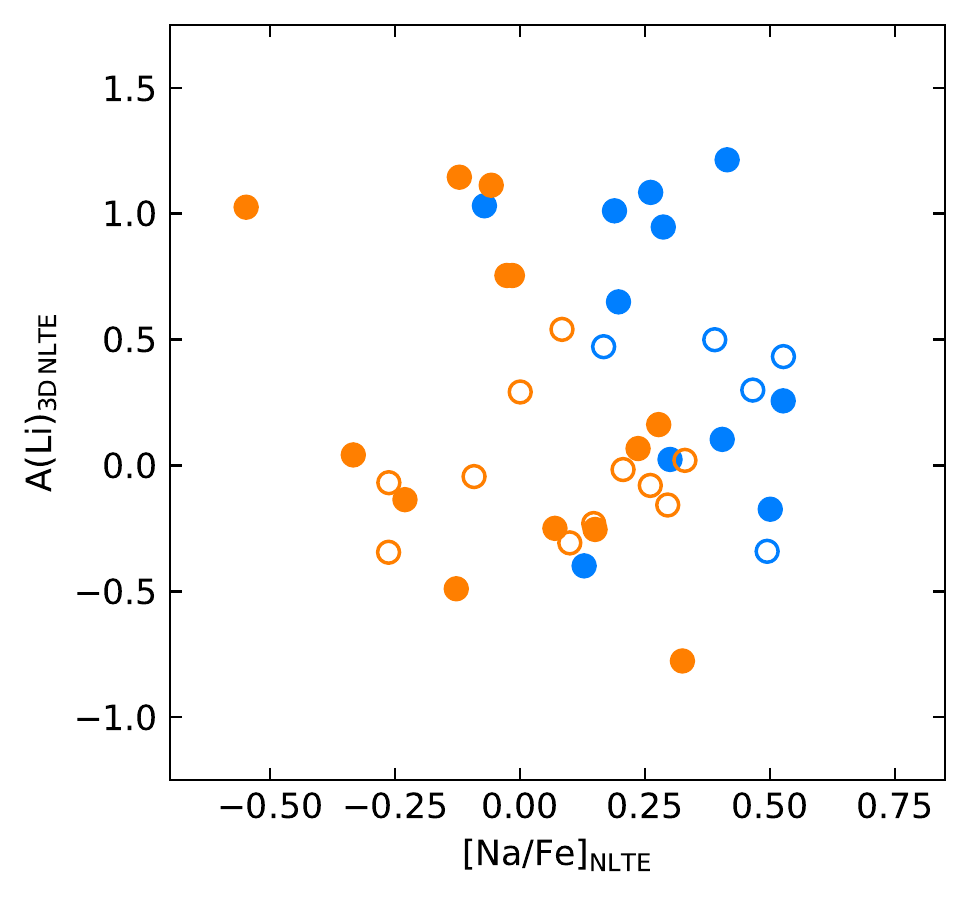}
 \caption{Lithium abundances in respect of [Na/Fe]. The meanings of symbols are the same as in Fig.~\ref{Fig12}.} 
\label{Fig14}
\end{figure}

The lithium abundances are not significantly larger in the metal-rich population of stars (see Figs.~\ref{Fig12} and \ref{Fig13}, where the upper values of abundances are shown by empty symbols). The difference in the averaged Li abundance values (Table~\ref{Table:4}) may be caused by a selection factor. Also, there is no correlation between the Li abundances and [Na/Fe] abundance ratios (Fig.~\ref{Fig14}). 
The $Gaia$-ESO Survey dataset of about a dozen globular clusters observed will enable the expansion of the statistical investigation
of the Li behaviour in globular clusters. In addition, it will foster further theoretical studies of stellar evolution and provide new insights into the multiple
stellar population problem in globular clusters (see the recent overview by \citealt{Sanna20}).
As yet, a complete understanding of the light-element
patterns and evolution has not been achieved (see \citealt{Randich21} for a review).

\begin{table*}
\caption{Main atmospheric parameters and elemental abundances in stars of the globular cluster NGC\,1851}
\label{Table:3}      %
\centering  
\small                                   
\begin{tabular*}{0.95\textwidth}{c c c c c c c c c c c c c c c}         
\hline\hline                    
Star    & $T_{\rm eff}$ & e\_$T_{\rm eff}$& log\,\textit{g}             & e\_log\,\textit{g}      &[Fe/H] &e\_[Fe/H] &$v_{\rm t}$ &e\_$v_{\rm t}$ &\multicolumn{3}{c}{Al\,{\sc i}} & ...& C/N \\    
                & K             &       K       &       &               & &&km\,s$^{-1}$  &km\,s$^{-1}$&[El/H] & $\sigma$ & nl & & & \\
\hline     
\noalign{\smallskip}                     
05141988$-$4003234      &4215   &       153     &       0.86    &       0.39    &       $-1.16$ &       0.12    &       1.74    &       0.09    &       $-1.03$ &       0.07    &       3       &       ...     &               \\
05135918$-$4002496      &4856   &       111     &       1.50    &       0.23    &       $-1.13$ &       0.09    &       1.82    &       0.11    &       $-0.74$ &       0.07    &       3       &       ...     &       0.72    \\
05141057$-$4003308      &4808   &       114     &       1.51    &       0.23    &       $-1.12$ &       0.10    &       1.69    &       0.04    &       $-0.89$ &       0.07    &       3       &       ...     &       0.58    \\
05134936$-$4003160      &3988   &       318     &       0.51    &       0.77    &       $-1.11$ &       0.18    &       2.16    &       0.05    &       $-0.69$ &       0.07    &       3       &       ...     &       0.78    \\
05141074$-$4004189      &4407   &       124     &       1.18    &       0.25    &       $-1.09$ &       0.10    &       1.59    &       0.03    &       $-0.98$ &       0.07    &       3       &       ...     &               \\
05141576$-$4003299      &4374   &       118     &       1.18    &       0.22    &       $-1.08$ &       0.10    &       1.56    &       0.05    &       $-0.84$ &       0.07    &       3       &       ...     &       0.65    \\
05133868$-$4007395      &4419   &       124     &       1.25    &       0.25    &       $-1.08$ &       0.10    &       1.66    &       0.04    &       $-0.91$ &       0.07    &       3       &       ...     &       0.91    \\
05140850$-$4005545      &4843   &       135     &       1.98    &       0.29    &       $-1.07$ &       0.11    &       1.45    &       0.03    &       $-0.47$ &       0.07    &       3       &       ...     &               \\
05135303$-$4000431      &4378   &       126     &       1.14    &       0.25    &       $-1.07$ &       0.11    &       1.63    &       0.04    &       $-0.95$ &       0.07    &       3       &       ...     &       0.81    \\
05141171$-$3959545      &4823   &       120     &       2.14    &       0.23    &       $-1.07$ &       0.10    &       1.43    &       0.03    &       $-0.53$ &       0.07    &       3       &       ...     &               \\
\hline                                             
\end{tabular*}
\tablefoot{We show here a part of Table~\ref{Table:3} to display the structure and content. The whole table can be found in electronic form at the CDS. It contains
the main atmospheric parameters and determined abundances for up to 28 chemical elements in 45 starts of NGC\,1851. "Star" -- the GES target identifier,  
"$T_{\rm eff}$" -- effective temperatures, "e\_$T_{\rm eff}$" -- uncertainty of the effective temperature, "log\,\textit{g}" -- surface gravity, 
"e\_log\,\textit{g}" -- uncerainty of surface gravity, "[Fe/H]" -- metallicity, "e\_[Fe/H]" -- uncertainty of metallicity,
"$v_{\rm t}$" -- microturbulent velocity, "e\_$v_{\rm t}$" -- uncertainty of microturbulent velocity, "[El/H]" -- element abundance in relation to hydrogen, "$\sigma$" -- abundance scatter between multiple lines for the same star, "nl" - number of lines used for the abundance determination.\\
}
\end{table*}

\begin{table}
\caption{Averaged abundances for the two populations and numbers of stars.}
\label{Table:4}
\centering
\begin{tabular}{lrrrrrr}
\hline\hline
\noalign{\smallskip}
\multirow{2}{*}{Parameter}      & \multicolumn{3}{c}{Metal-poor} & \multicolumn{3}{c}{Metal-rich} \\
        &   Average    &   $\sigma$ & N &  Average &  $\sigma$ & N \\
\hline
\noalign{\smallskip}
\,[Fe/H] & $-1.05$&0.05 &28 &$-0.98$ &0.04 &17 \\
\,A(Li\,{\sc i})\,$_{\rm 3D\,NLTE}$ &   0.14 &  0.66 &  27 &   0.46 &  0.59 &  17 \\
\noalign{\smallskip}
\,[C/Fe]\,(C$_2$)   &  $-0.31$ &  0.08 &  14 &  $-0.36$ &  0.11 &   9 \\
\,[N/Fe]\,(CN)   &   0.47 &  0.09 &  12 &   0.73 &  0.14 &   9 \\
\,[O/Fe]\,([O\,{\sc i}])  &   0.27 &  0.18 &  28 &  $-0.13$ &  0.20 &  16 \\
\,[Na\,{\sc i}/Fe]\,$_{\rm NLTE}$ &   0.02 &  0.23 &  25 &   0.32 &  0.17 &  16 \\
\,[Mg\,{\sc i}/Fe]  &   0.26 &  0.08 &  28 &   0.25 &  0.10 &  17 \\
\,[Al\,{\sc i}/Fe]  &   0.17 &  0.19 &  28 &   0.38 &  0.15 &  17 \\
\,[Si\,{\sc i}/Fe]  &   0.07 &  0.05 &  28 &   0.09 &  0.05 &  17 \\
\,[Ca\,{\sc i}/Fe]  &   0.17 &  0.05 &  28 &   0.21 &  0.05 &  17 \\
\,[Ca\,{\sc ii}/Fe] &   0.24 &  0.12 &  25 &   0.29 &  0.15 &  15 \\
\,[Ti\,{\sc i}/Fe]  &   0.14 &  0.07 &  28 &   0.16 &  0.08 &  17 \\
\,[Ti\,{\sc ii}/Fe] &   0.17 &  0.05 &  28 &   0.18 &  0.04 &  17 \\
\,[Sc\,{\sc i}/Fe] &  $-0.01$ &  0.12 &  28 &  $-0.02$ &  0.12 &  17 \\
\,[Sc\,{\sc ii}/Fe]  &  $-0.01$ &  0.04 &  28 &  $-0.02$ &  0.05 &  17 \\
\,[V\,{\sc i}/Fe]   &  $-0.11$ &  0.07 &  28 &  $-0.08$ &  0.08 &  17 \\
\,[Cr\,{\sc i}/Fe]  &  $-0.16$ &  0.06 &  28 &  $-0.12$ &  0.07 &  17 \\
\,[Cr\,{\sc ii}/Fe] &   0.04 &  0.11 &  27 &  $-0.05$ &  0.08 &  17 \\
\,[Mn\,{\sc i}/Fe] &  $-0.37$ &  0.05 &  28 &  $-0.41$ &  0.05 &  17 \\
\,[Co\,{\sc i}/Fe]  &  $-0.09$ &  0.04 &  28 &  $-0.07$ &  0.04 &  17 \\
\,[Ni\,{\sc i}/Fe]  &  $-0.14$ &  0.04 &  28 &  $-0.15$ &  0.06 &  17 \\
\,[Cu\,{\sc i}/Fe]  &  $-0.33$ &  0.17 &  28 &  $-0.28$ &  0.16 &  17 \\
\,[Zn\,{\sc i}/Fe]  &  $-0.06$ &  0.10 &  28 &  0.00 &  0.15 &  17 \\
\,[Y\,{\sc ii}/Fe]   &  $-0.24$ &  0.06 &  28 &  $-0.02$ &  0.14 &  17 \\
\,[Zr\,{\sc i}/Fe]  &   0.15 &  0.18 &  27 &   0.37 &  0.15 &  17 \\
\,[Mo\,{\sc i}/Fe]  &   0.16 &  0.16 &  14 &   0.37 &  0.13 &  12 \\
\,[Ba\,{\sc ii}/Fe]  &   0.10 &  0.13 &  28 &   0.55 &  0.19 &  17 \\
\,[Ba\,{\sc ii}/Fe]\,$_{\rm NLTE}$ &  0.00 &  0.15 &  28 &   0.47 &  0.20 &  17 \\
\,[La\,{\sc ii}/Fe]  &  $-0.03$ &  0.14 &  27 &   0.37 &  0.10 &  17 \\
\,[Ce\,{\sc ii}/Fe]  &  $-0.18$ &  0.15 &  28 &   0.32 &  0.19 &  17 \\
\,[Pr\,{\sc ii}/Fe]  &   0.46 &  0.05 &  15 &   0.63 &  0.12 &  12 \\
\,[Nd\,{\sc ii}/Fe]  &   0.27 &  0.07 &  28 &   0.52 &  0.14 &  17 \\
\,[Sm\,{\sc ii}/Fe]  &   0.37 &  0.13 &  14 &   0.55 &  0.15 &  12 \\
\,[Eu\,{\sc ii}/Fe]  &   0.40 &  0.10 &  28 &   0.41 &  0.07 &  17 \\
\noalign{\smallskip}
\,A(C+N+O)            &   7.97 &  0.11 &  12 &   7.94 &  0.08 &   9 \\
\,C/N              &   0.66 &  0.13 &  12 &   0.35 &  0.13 &   9 \\
\,[$\alpha$/Fe]         &   0.16 &  0.04 &  28 &   0.18 &  0.05 &  17 \\
\,[Iron peak/Fe]      &  -0.09 &  0.04 &  28 &  -0.07 &  0.04 &  17 \\
\,[$s$/Fe] &  $-0.01$ &  0.08 &  28 &   0.34 &  0.11 &  17 \\
\hline                                                                                                                                  
\end{tabular}
\end{table}

\section{Formation scenario for NGC\,1851}

\subsection{A hint of different ages and the two-cluster merger based on C, N, and O abundances }

There have been a number of studies that found that strengths of CN bands vary in stars of NGC\,1851 quite significantly (e.g. \citealt{Pancino10, Lardo12}). Moreover, \citet{Campbell12} and \citet{Simpson17} updated the picture by finding in low-resolution spectra of red giants and asymptotic giant branch stars a quadrimodal distribution of CN band strengths, which could be explained by a superposition of two bimodal populations of two merged clusters. Our results also show spreads of the C/N ratios in both subsamples.  
As in our study, low-resolution spectral investigations of larger samples of the cluster members exhibited the multi-modal behaviour in their CN, but not in their CH bands (\citealt{Simpson17, Lim15}). 
The high-precision photometric study of \citet{Milone08} interpreted the two distinct subgiant branches in NGC\,1851 as evidence for some combination of bimodalities in age and total [(C+N+O)/Fe] abundance. This latter possibility was supported by \citet{Yong09, Yong15}, who in their study of 11 stars, found a difference of $0.6$~dex in [(C+N+O)/Fe] between the two populations. A smaller spread in $A$(C+N+O) was obtained by \citet{Simpson17} using medium-resolution spectra. \citet{Villanova10} investigated high-resolution spectra of 15 RGB stars and, on the contrary, found that the two  populations do not show any significant difference in their total $A$(C+N+O) content. 

\begin{figure}
\centering
 \includegraphics[width=0.45\textwidth]{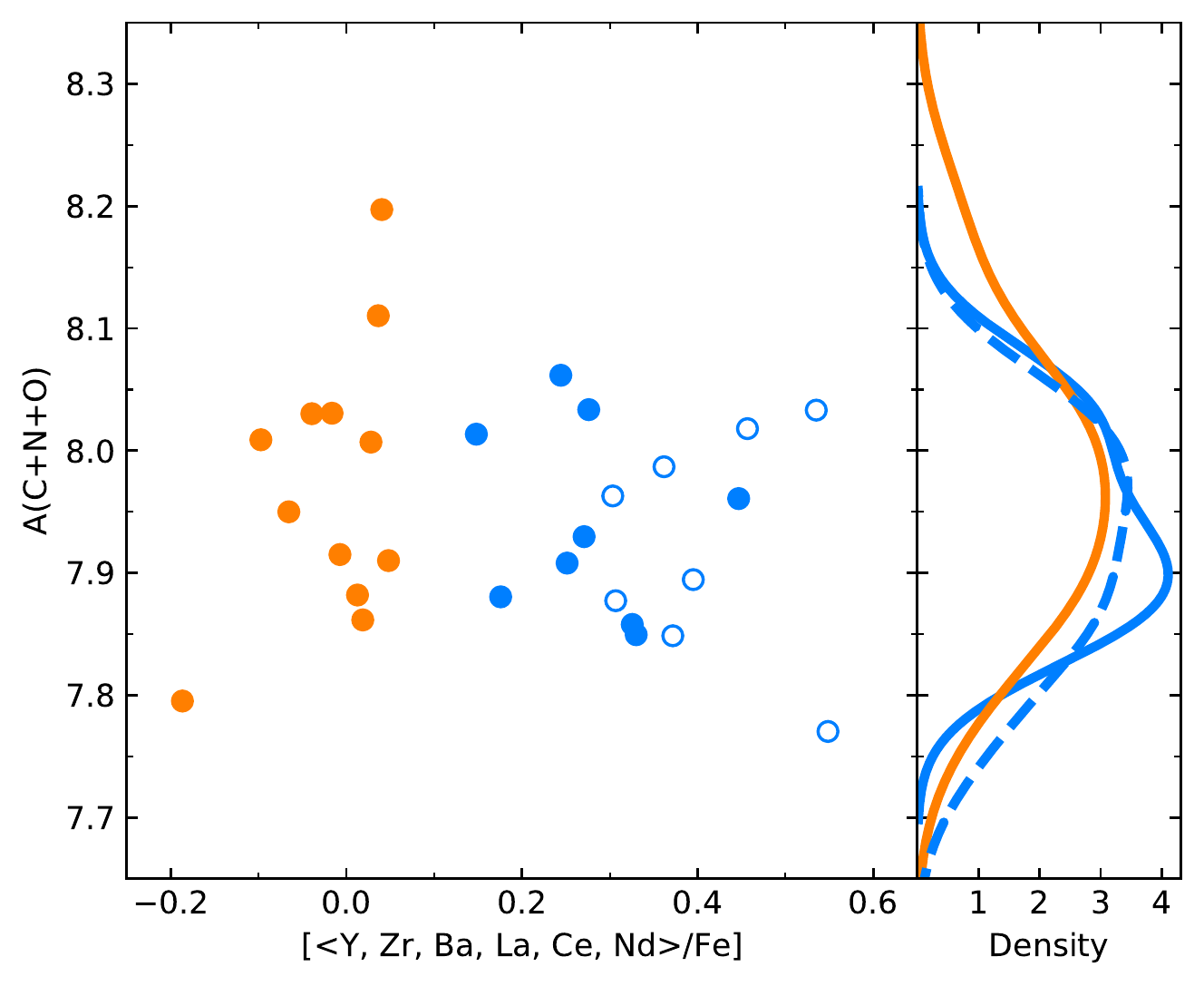}
 \caption{$A$(C+N+O) abundances in respect to the averaged abundances of the $s$-process dominated chemical elements. The dashed blue line shows the density with the values marked by the empty symbols. The meaning of symbols is as in Fig.~\ref{Fig8}. The $A$(C+N+O) abundances are similar in both populations of NGC\,1851 stars. } 
\label{Fig15}
\end{figure}

Our study shows that there is a spread of about $\pm0.1$~dex in $A$(C+N+O) in both populations, however, the averaged values between the populations do not differ (see Fig.~\ref{Fig15} and Table~\ref{Table:4}).  
\citet{Gratton12} evaluated that the metal-rich subsample would be older by $\sim0.6$~Gyr than the metal-poor if the total $A$(C+N+O) abundance is the same in both subsamples (see their Fig.~20). In addition, for instance, if the sum of $A$(C+N+O) were larger by a factor of 2 for metal-rich stars than for metal-poor stars, then the first would
be found to be younger than the second by $\sim$\,0.4~Gyr; however, the actual values are
almost fully inverted. 

Thus, our robust analysis of C, N, and O abundances provides additional evidence that NGC\,1851 is composed of two clusters, since the metal-rich cluster is by about 0.6~Gyr older than the metal-poor one. The recent overall age estimate for NGC\,1851 is $12.27 (+1.47/-0.98)$~Gyr (\citealt{Valcin20}). 

The older and metal-rich cluster has a different chemical composition pattern than the metal-poor one and might be formed from material enriched by ejecta of asymptotic giant branch stars or other polluters (e.g. \citealt{deMink09, Bastian13, Bastian15, Gieles18}). 

While the possibility for globular clusters to merge in the Galactic halo seems unrealistic, this kind of merger can happen in dwarf galaxies. Numerical simulations by \citet{Bekki12} showed that two clusters can merge and form the nuclear star cluster of a dwarf galaxy. After the parent dwarf galaxy is accreted by the Milky Way, its dark matter halo and stellar envelope can be stripped by the Galactic tidal field, leaving behind the nucleus (i.e. NGC 1851) and a diffuse stellar halo. 

\begin{figure}
\centering
 \includegraphics[width=0.48\textwidth]{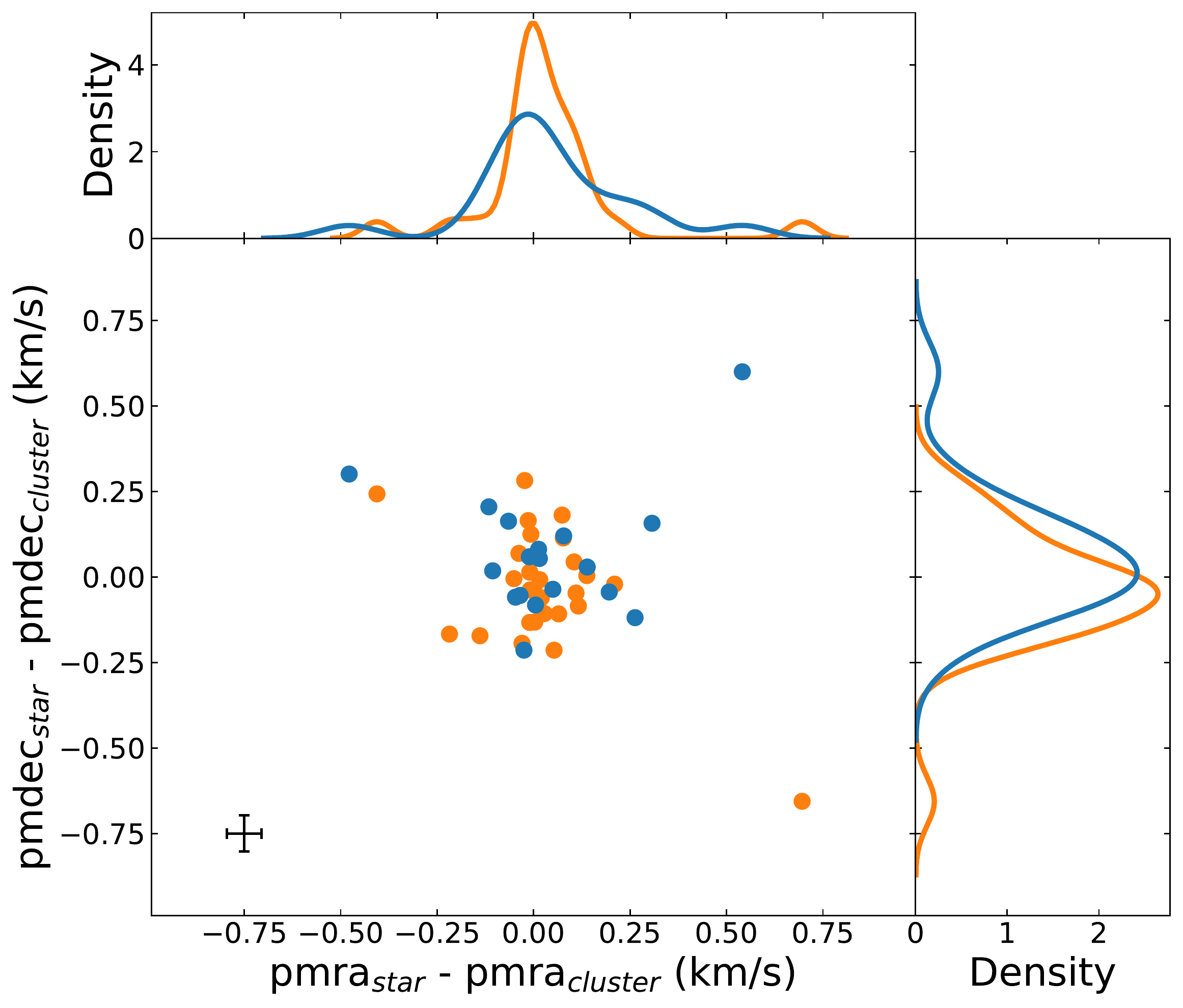}
 \caption{Distribution of proper motions of stars in respect to the mean values of the cluster. The proper motions of the cluster stars were taken from the $2^{\rm nd}$ $Gaia$ data release \citep{GaiaDR2_2018}. The meaning of symbols is as in Fig.~\ref{Fig4}.}  
\label{Fig16}
\end{figure}

In principle, stars from the two merged clusters could also have different kinematics, which would then be visible in their velocity distribution. We selected stellar proper motions measured by the $Gaia$ space telescope \citep{Gaia2016, GaiaDR2_2018} and compared the metal-poor and metal-rich subsamples in Fig.~\ref{Fig16}. 
As in \citet{Carretta11}, we found no statistically significant differences. It would be good to have a comprehensive dynamical model, however, it is not known when the merging occurred and how the relaxation proceeded. If some
debris from the parent galaxy of NGC\,1851  still exists (as discussed in the next section),  then the event could have taken place rather recently and the cluster is not yet relaxed.  

\subsection{Parent galaxy}

A possibility that NGC\,1851 originated at the nucleus of a dwarf galaxy that was captured by the Milky Way was supported by the work of \citet{Marino14}, who confirmed the presence of a diffuse `halo' of stars co-moving with the cluster (noted by \citealt{Olszewski09}) and found that it has the same chemical pattern as the $s$-process-poor stars within the cluster. Fifteen stars were interpreted as being part of the field population from the original dwarf galaxy. \citet{Navin15} also found four probable extratidal cluster halo stars at distances up to $\sim3.1$ times the tidal radius (confirmed by \citealt{Simpson17}), which support the notion that NGC\,1851 is surrounded by an extended stellar halo. \citet{Kuzma18} confirmed the existence of a large
diffuse stellar envelope surrounding NGC\,1851 of size at least 240~pc in radius.
According to results of the RAVE survey reported by \citet{Kunder14}, stars that may be associated with NGC\,1851 reach projected distances of about 10 degrees away from the core of the cluster. 
These findings are rather reliable as the radial velocity and [Fe/H] abundance of NGC\,1851 are largely offset from the Galactic disc field stars seen in projection along the line of sight; thus, the field star contamination issues can be easily resolved.

\begin{figure*}
 \includegraphics[width=1.00\textwidth]{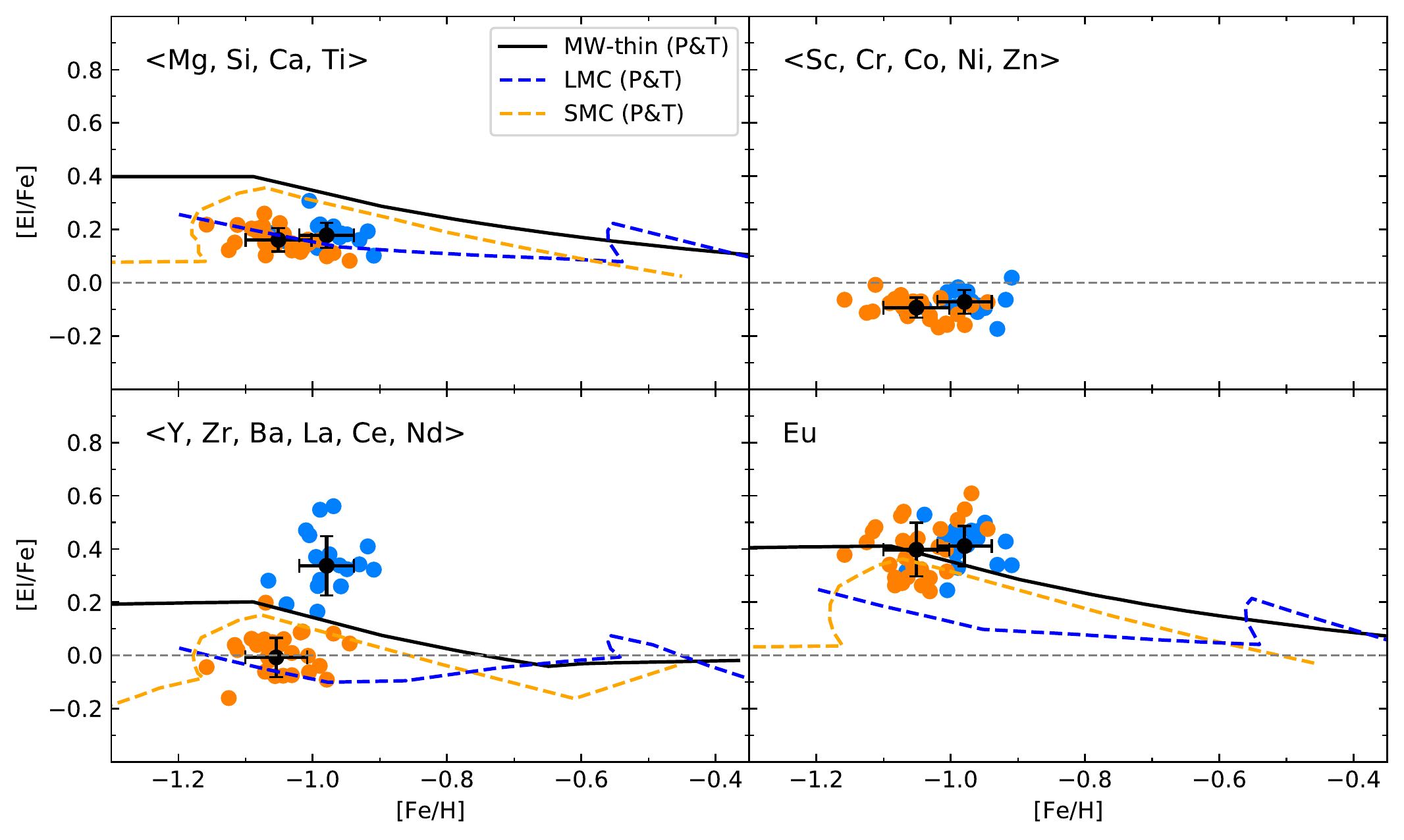}
 \caption{Comparison of mean abundances of chemical elements dominated by different nucleosynthesis processes with the corresponding models of the Milky Way and the Large and Small Magellanic Cloud evolution by \citet{Pagel95, Pagel97, Pagel98}. 
 The meaning of other symbols is as in Fig.~\ref{Fig5}.  
} 
\label{Fig17}
 \end{figure*}

\citet{Marino14} noted that NGC\,1851 lies within the so-called `Vast Polar Structure' of satellite galaxies of the Milky Way discussed in \citet{Pawlowski13} and references therein. The majority of satellite galaxies of the Milky Way define a thin plane perpendicular to the Galactic disc. 
Recently, based on the Early Gaia Data Release, \citet{Li21} estimated proper motions of 46 dwarf galaxies of the Milky Way and found that between half and two-thirds of the satellites have orbital poles that indicate them to orbit along with the vast polar structure, with the vast majority of these co-orbiting in a common direction also shared by the Magellanic Clouds. This is indicative of the real structure of dwarf galaxies. 

NGC\,1851 was often compared with other globular clusters that have a spread of heavy chemical elements in their stars, such as M\,22 or $\omega$\,Centauri. Even though some similarities have been revealed, the formation and evolution histories of these star clusters were quite different. As an example, \citet{Lee20} found five populations in M\,22 from Ca-CN-CH photometry. The metal-poor stars were divided into two subpopulations and the metal-rich stars into three subpopulations. They support the idea that M\,22 was formed via a merger of two globular clusters. \citet{Meszaros21} investigated multiple populations of $\omega$\,Centauri by analysing APOGEE spectra and found seven populations based on stellar [Fe/H], [Al/Fe], and [Mg/Fe] abundances, as well as the increased $A$(C+N+O) with increased metallicity. The parent galaxies of these clusters also may be different.

Key data is derived from kinematics and ages. 
There is a number of recent studies based on the new $Gaia$ space mission results that investigate what merger the peculiar NGC\,1851 cluster and other clusters could be attributed to. 
For instance, \citet{Massari19} attributed NGC\,1851 to the Gaia-Enceladus (along with 25 other globular clusters and 6 tentative ones), while M\,22 was attributed to the Galactic main disc, and $\omega$\,Centauri was attributed to the Gaia-Enceladus or Sequoia merger event  (\citealt{Forbes20} is in favour of $\omega$\,Centauri rather belonging to Sequoia).

According to \citet{Helmi18}, the Gaia-Enceladus was slightly more massive than the Small Magellanic Cloud. 
In Fig.~\ref{Fig17}, for the metal-rich and metal-poor NGC\,1851 subsamples,
we displayed the mean abundances of chemical elements
dominated by different nucleosynthesis processes  and compared with
the corresponding semi-empirical models of the Large and Small Magellanic
Clouds from \citet{Pagel98} and to the Milky Way
models from \citet{Pagel95, Pagel97}. 
It can be seen that NGC\,1851 $\alpha$-elemental abundances are lower than in the Milky Way
 and are very similar to those in the Large Magellanic Cloud, as well as to the model predictions at this metallicity compiled for the Gaia-Enceladus galaxy (see Fig.~2 \citealt{Vincenzo19}).    Abundances of the $r$-process dominated element europium are larger, meaning that in the progenitor $r$-process, elements were 
produced more efficiently relative to $\alpha$-process elements. The abundance ratios of $s$-process dominated elements to iron in the metal-poor subsample are similar to the Large Magellanic Cloud. A similar pattern was found by \citep{Matsuno21} in a sample of 47 stars attributed to the Gaia-Enceladus in the GALAH survey (abundances of Mg, Fe, Ba, La, and Eu were determined). The authors also performed one-zone chemical evolution calculations and showed that high [Eu/Mg]
and low [Mg/Fe] ratios can be explained by chemical enrichment from neutron star mergers and Type~Ia supernovae. The elevated $r$-process element enrichment in four Gaia-Enceladus and five Sequoia stars was determined also by \citet{Aguado21}, as well as in two stars of the globular cluster NGC\,1261 (associated with the Gaia-Enceladus) by \citet{Koch21}. Thus, according to its chemistry, NGC\,1851 fits the Gaia-Enceladus parent galaxy and even may be its nucleus as suggested by \citet{Bekki12} and \citet{Forbes20}.    

\section{Summary and conclusions}

In this work, we investigated abundances of 29 chemical elements in 45  giants of the globular cluster NGC\,1851 using high-resolution spectra within the $Gaia$-ESO Public Spectroscopic Survey. The stars were separated to metal-rich and metal-poor subsamples   according  to  their  abundances  of  nitrogen  and $s$-process dominated elements. Special attention was given to establishing a robust determination of C, N, and O abundances. Our results can be summarised as follows: 

\begin{itemize}
\item

The investigated stars can be separated into two subsamples with a difference of 0.07~dex in the mean metallicity. There are 17 metal-rich and 28 metal-poor stars, with averaged metallicities of $-0.98\pm0.04$~dex and $-1.05\pm0.05$~dex, respectively.

\item
The subsamples have a difference of 0.3 in the mean C/N and 0.35~dex in the mean [$s$/Fe] values. 

\item
The two subsamples also differ in the mean abundances of oxygen, sodium, and aluminium, and both have Na-O anticorrelations.

\item
No significant difference was determined in the mean abundance to iron ratios of lithium, carbon, $\alpha$- and iron-peak-elements, and of europium. 
 
\item
There is a spread of about $\pm 0.1$~dex in $A$(C+N+O) in  both  subsamples,  however,  the  averaged  values between the subsamples do not differ, which provides additional evidence that NGC\,1851 is composed of two clusters, with the metal-rich cluster being by about 0.6~Gyr older than the metal-poor one.

\item
A global overview of NGC\,1851 properties and the detailed abundances of chemical elements favour a formation scenario in a dwarf spheroidal galaxy accreted by the Milky Way. 

\end{itemize}

In order to understand the complex history of  the origins of globular clusters in the Milky Way, many more comprehensive studies will have to be accomplished.   

\begin{acknowledgements}
Based on the Gaia-ESO Public Spectroscopic Survey data products from observations made with the ESO Very Large Telescope at the La Silla Paranal Observatory under programme ID 188.B-3002. These data products have been processed by the Cambridge Astronomy Survey Unit (CASU) at the Institute of Astronomy, University of Cambridge, and the FLAMES/UVES reduction team at INAF/Osservatorio Astrofisico di Arcetri. The Gaia-ESO Survey Data Archive is prepared and hosted by the Wide Field Astronomy Unit, Institute for Astronomy, University of Edinburgh, which is funded by the UK Science and Technology Facilities Council. 
The anonymous referee is thanked for helpful suggestions.
S.L.M acknowledges support from the Australian Research Council through grant DP180101791 and from the UNSW Scientia Fellowship program. 
T.B. was funded by grant No. 2018-04857 from The Swedish Research Council.
U.H. acknowledges support from the Swedish National Space Agency (SNSA/Rymdstyrelsen).
This research has made use of SIMBAD (operated at CDS, Strasbourg), of VALD (\citealt{Kupka00}), and of NASA’s Astrophysics Data System.  
This work has made use of data from the European Space Agency (ESA) mission
{\it Gaia} (\url{https://www.cosmos.esa.int/gaia}), processed by the {\it Gaia}
Data Processing and Analysis Consortium (DPAC,
\url{https://www.cosmos.esa.int/web/gaia/dpac/consortium}). Funding for the DPAC
has been provided by national institutions, in particular the institutions
participating in the {\it Gaia} Multilateral Agreement.

\end{acknowledgements}

\bibliographystyle{aa} 
\bibliography{References3.bib} 

\end{document}